\documentclass[5p, final, twocolumn, 10pt, a4paper]{elsarticle}
\biboptions{sort&compress}

\usepackage[utf8]{inputenc}
\usepackage{amsmath,amssymb,amsfonts}
\usepackage{algorithmic}
\usepackage{graphicx}
\usepackage{grffile}
\usepackage{textcomp}
\usepackage[caption=false,font=footnotesize]{subfig}
\usepackage[hidelinks]{hyperref}
\usepackage{balance}
\usepackage{bm}
\usepackage{tikz}
\usetikzlibrary{shapes, arrows, positioning, calc}
\usepackage{circuitikz}
\usepackage{pgfplots}
\usepackage{stfloats}
\usepackage{enumitem} 
\usepackage{microtype}
\usepackage{everypage}

\pgfplotsset{compat=1.15}

\newcommand{\suma}{\Large$+$}

\newcommand{\mult}{\Large$\times$}

\tikzset{%
  block/.style    = {draw, thick, rectangle, minimum height = 3em,
    minimum width = 3em},
  sum/.style      = {draw, circle, node distance = 2cm}, 
  input/.style    = {coordinate}, 
  output/.style   = {coordinate} 
}

\graphicspath{ {./images/} }

\journal{Signal Processing: Image Communication}

\makeatletter
\def\ps@pprintTitle{%
\let\@oddhead\@empty
\let\@evenhead\@empty
\def\@oddfoot{\centerline{\thepage}}%
\let\@evenfoot\@oddfoot}
\makeatother

\begin{document}

\begin{frontmatter}



\title{Single-pixel compressive imaging in shift-invariant spaces via exact wavelet frames}


\author{Tin Vla\v{s}i\'{c}\corref{cor1}}
\cortext[cor1]{Corresponding author.}
\ead{tin.vlasic@fer.hr}

\author{Damir Ser\v{s}i\'{c}\corref{}}
\ead{damir.sersic@fer.hr}

\address{Department of Electronic Systems and Information Processing, University of Zagreb Faculty of Electrical Engineering and Computing, Unska 3, HR-10000 Zagreb, Croatia}

\begin{abstract}
This paper introduces a novel framework for single-pixel imaging via compressive sensing (CS) in shift-invariant (SI) spaces by exploiting the sparsity property of a wavelet representation. We reinterpret the acquisition procedure of a single-pixel camera as filtering of the observed signal with continuous-domain functions that lie in an SI subspace spanned by the integer shifts of the box function. The signal is modeled by an arbitrary SI generator whose special case is the box function, which, as we show in the paper, is conventionally used in single-pixel imaging. We propose to use separable B-spline generators which are intuitively complemented by sparsity-inducing spline wavelets. The SI models of the acquisition and the underlying signal lead to an exact discretization of an inherently continuous-domain inverse problem to a finite-dimensional problem of CS type. By solving the CS optimization problem, a parametric representation of the signal is obtained. Such a representation offers many practical advantages in image processing applications. We propose an efficient matrix-free implementation of the framework and conduct it on the standard test images and real-world measurement data. Experimental results show that the proposed framework achieves a significant improvement of the reconstruction quality relative to the conventional discretization in CS setups. MATLAB implementation of the method described in this paper has been made publicly available on \url{https://github.com/retiro/compressive_imaging_in_si_spaces}.
\end{abstract}


\end{frontmatter}

\thispagestyle{empty}

\section{Introduction} \label{sec:Intro}
Sampling theories are concerned with the reconstruction of analog signals from discrete samples. Being an ill-posed inverse problem, reconstruction from samples must rely on some signal prior, \textit{e.g.}, an assumption that the observed signal lies in a predefined subspace. In conventional sampling, famously embodied in the Nyquist-Shannon theorem \cite{Nyquist:SNTheorem, Shannon:SNTheorem}, signals are assumed to lie in a shift-invariant (SI) subspace spanned by the integer shifts of the $\mathrm{sinc}$ function, where ${\mathrm{sinc}(t)=\sin(\pi t)/(\pi t)}$. The Nyquist-Shannon theorem was generalized to a wider class of SI subspaces \cite{Unser:GeneralSampTheory, Unser:50YearsAfterShannon, Eldar:BeyondBandlimitedSampling}, which allows more realistic sampling schemes and signal representations, including splines \cite{Unser:APerfectFit} and wavelets \cite{Mallat:MRA, Daubechies:Wavelets}. To obtain an exact reconstruction, the signal has to be sampled at least at the rate of innovation \cite{Vetterli:FRI}, which corresponds to the Nyquist limit in the conventional sampling theorem.

During the past two decades, sparsity has become an important signal prior in the field of signal processing. It lies at the heart of compressive sensing (CS) \cite{Candes:CS, Donoho:CS, Foucart:IntroCS}, a sampling and reconstruction method that goes beyond the Nyquist limit. In standard CS, an unknown signal ${f:D \rightarrow \mathbb{R}}$, where $D$ is a region of interest, is assumed to be exactly represented as the linear combination ${\sum_{n=1}^N x_n \xi_n}$ of a finite number of basis functions. The coefficients $\{x_n\}_{n=1}^N$ are a discrete representation of the observed continuous-domain signal and are assumed to be $Q$-sparse, where ${Q \ll N}$. The measurement procedure is described by a linear operator $\bm{z}$, which yields discrete samples ${\bm{y}=\bm{z}(f) \in \mathbb{R}^M}$, where ${2Q \lesssim M<N}$. Signal reconstruction is then given by the optimization problem
\begin{equation}
    \label{eq:Intro_CsProblem}
    \arg \min_{\bm{x}\in \mathbb{R}^N} \Vert \bm{y} - \bm{\Theta x} \Vert_{\ell_2}^2 + \lambda \Vert \bm{x} \Vert_{\ell_1},
\end{equation}
where ${\bm{\Theta}:\mathbb{R}^N \rightarrow \mathbb{R}^M}$ is a discrete version of the forward model, with elements ${[\bm{\Theta}]_{m,n}=\langle z_m, \xi_n \rangle}$. While the $\ell_2$ norm is used to measure the consistency of the solution with the measurements, the $\ell_1$-norm regularization promotes sparse solutions \cite{Donoho:L1Sparsest}. The regularization parameter $\lambda$ in \eqref{eq:Intro_CsProblem} sets a tradeoff between sparsity and $\ell_2$-norm error. A faithful signal recovery from ${M\gtrsim 2Q}$ linear measurements is possible under strict conditions on $\bm{\Theta}$, namely \textit{restricted isometry property} (RIP) and \textit{incoherence} \cite{Candes:CS, Candes:RIP, Roman:2014asymptotic, Adcock:BreakingCoherence}.

In the vast majority of papers, the discrete version of a continuous-domain forward model is an approximation that introduces errors to real-world implementations of CS. Furthermore, one typically assumes the bandlimited model of the underlying signal, which may additionally contribute to the reconstruction error if its continuous-domain generator is not considered in the discretization procedure. There have been several attempts to generalize CS to infinite-dimensional problems, that were reported in \cite{Mishali:Xampling, Adcock:GenSamplAndInftyCS, Adcock:InftyDimCS}. The SI signal model is employed in a few works to exactly discretize continuous-domain inverse problems into CS-type problems. In \cite{Eldar:CSAnalogSI} and \cite{Eldar:RobustRecovery}, Eldar and Mishali use CS for reconstruction of signals in unions of SI subspaces. Sparsity is modeled by assuming that only a few out of all generators in the union are active. The authors of  \cite{Unser:SplinesUnSolution, Debarre2019, Debarre:HybridDictionaries, Bohra2020} use splines as a representation basis of the underlying signal for the discretization of the continuous-domain inverse problems. The signal is assumed to be sparse in that it has only a few innovations per time unit. A different approach is considered in \cite{Vlasic:EUSIPCO} and \cite{Vlasic:CSinSIspaces}, where an SI model of the observed one-dimensional signal is used to exactly discretize an inherently continuous-time inverse problem, and the expansion coefficients of the SI model are assumed to be sparse in a discrete transform domain.

In this paper, we propose to use SI models for the discretization of the observed signal and the acquisition procedure in single-pixel imaging. Over the past decade, single-pixel imaging has been used in many applications and has several areas of possible advantage \cite{Duarte:SinglePxCamera, Edgar2019, Gibson2020}. A single-pixel camera consists of two components - the light modulator and the single-pixel detector. The light modulator is usually implemented by the digital micromirror device (DMD), which selectively redirects the light beams to the photodetector or projects the sequence of light patterns onto the object that is being captured by the detector. We model the DMD modulation procedure with two-dimensional filtering by separable functions that lie in an SI subspace spanned by the box function. This allows us to model the observed underlying signal in a more appropriate continuous-domain SI space than in the conventional pixel-by-pixel basis which we argue that is too coarse an approximation. Consequently, we are able to link the theories of generalized sampling in SI spaces and CS to discretize the inherently continuous-domain inverse problem of single-pixel imaging exactly. We propose to use polynomial B-splines as a representation basis of the signal. The B-splines lead to an efficient implementation of the proposed framework and provide many practical advantages in image processing once we recover their expansion coefficients directly from a reduced set of measurements. We employ spline wavelets as sparsity-inducing bases, which offer an intuitive complement of the B-spline function spaces. The reconstruction quality of the framework was assessed on the standard test images and data acquired by an off-the-shelf realization of the single-pixel camera. Experimental results show that the proposed framework improves the reconstruction quality relative to single-pixel imaging with the conventional discretization technique.

The main contributions of the paper are as follows:
\begin{itemize}[noitemsep, nolistsep]
    \item Reinterpretation of the single-pixel camera's acquisition procedure as filtering of the continuous-domain signal by a sampling signal that lies in an SI subspace spanned by the two-dimensional box functions;
    \item Continuous-domain approach to single-pixel compressive imaging which elegantly links the theories of generalized sampling, multiresolution representation and CS, and results in an exact discretization of the inherently continuous-domain inverse problem;
    \item Matrix-free implementation of the proposed method which fits with the memory efficient solvers for CS;
    \item Extensive experiments of the proposed method conducted on the test images and real-world measurement data, which show a significant improvement in comparison to the conventional discretization method.
\end{itemize}

The paper is organized as follows: we give a short reminder on multiresolution representation and link with the generalized sampling in SI spaces in Section \ref{sec:background}. The emphasis is put on the orthogonal projection of signals onto the Riesz bases, which is very important for the understanding of the manuscript in the following sections. In Section \ref{sec:CSinSIspaces}, we propose a novel framework for single-pixel imaging via CS in SI spaces based on a wavelet-domain sparsity assumption. In Section \ref{sec:Implementation}, an efficient implementation of the proposed framework for large-scale images is suggested. Experimental results are illustrated in Section \ref{sec:Experiments}. We relate our work to the findings of other studies in Section \ref{sec:RelatedWork} and conclude the paper in Section \ref{sec:Conclusion}.

\section{Background: wavelets and generalized sampling}
\label{sec:background}
We deal with the problem of recovering and representing an unknown function $f(t)$ that lies in the Hilbert space $L_2$, which consists of functions that are square-integrable and whose inner product is defined by
\begin{equation}
    \label{eq:L2innerProduct}
    \langle f(t), g(t) \rangle = \int \limits_{-\infty}^{\infty} f^{*}(t)g(t)dt,
\end{equation}
for some ${g(t) \in L_2}$. In order to recover $f(t)$ from its uniform samples, it is commonly assumed that $f(t)$ lies in an appropriate SI subspace $\mathcal{V}_{(0)}$ of $L_2$ \cite{Nyquist:SNTheorem, Shannon:SNTheorem, Unser:GeneralSampTheory, Unser:50YearsAfterShannon, Eldar:BeyondBandlimitedSampling}. Any signal ${f(t) \in \mathcal{V}_{(0)}}$ has the form
\begin{equation}
    \label{eq:SignalinSIspace}
    f(t) = \sum \limits_{k \in \mathbb{Z}} a_{(0)}[k] \varphi(t-kT),
\end{equation}
for some subspace generator $\varphi(t)$ and sampling period $T$. For simplicity, throughout the manuscript, we set ${T=1}$.

To guarantee a unique and stable representation of any signal in $\mathcal{V}_{(0)}$ by expansion coefficients $a_{(0)}[k]$, the generator should form a Riesz basis for $L_2$ \cite{Unser:50YearsAfterShannon, Eldar:BeyondBandlimitedSampling}. A countable set of vectors ${\{ \varphi(t-k) \}}$ in $L_2$ is a Riesz basis for $L_2$ if it is complete and if and only if there exist two positive constants ${\alpha > 0}$ and ${\beta<\infty}$ such that
\begin{equation}
    \label{eq:RieszBasisCondition}
    \alpha\left\Vert \bm{a}_{(0)} \right\Vert_{\ell_2}^{2} \leq \left\Vert \sum \limits_{k \in \mathbb{Z}} a_{(0)}[k] \varphi(t-k) \right\Vert^2_{L_2} \leq \beta \left\Vert \bm{a}_{(0)} \right\Vert_{\ell_2}^{2},
\end{equation}
where ${\Vert \bm{a}_{(0)} \Vert_{\ell_2}^{2} = \sum_k\vert a_{(0)}[k] \vert^2}$ is the squared $\ell_2$ norm of the expansion coefficients $a_{(0)}[k]$. Namely, a Riesz basis is a set of linearly independent vectors which ensures that a small modification of the expansion coefficients results in a small distortion of the signal representation. In particular, the basis is orthonormal if and only if ${\alpha = \beta = 1}$. Notice that a set ${\{ \varphi(t-k) \}}$ in $L_2$ is an exact frame for $L_2$ if it is a Riesz basis for $L_2$ \cite{Christensen2003}.

\subsection{Multiresolution representation}
\label{subsec:MultiresRepresentation}
We assume that $\mathcal{V}_{(0)}$ is a subspace in a sequence of nested subspaces such that ${\mathcal{V}_{(i)} \supset \mathcal{V}_{(i+1)}}$ for ${i \in \mathbb{Z}}$. These subspaces are spanned by a set of Riesz basis vectors ${\{ \varphi(2^{-i}t-k) \}}$, obtained by dilation and translation of a generator $\varphi(t)$ \cite{Mallat:MRA, Daubechies:Wavelets}, which is also referred to as a scaling function. Approximation of ${f(t) \in L_2}$ at resolution $2^{i}$ in $\mathcal{V}_{(i)} $ is defined by
\begin{equation*}
    \label{eq:AppInNestedSubspaces}
    \hat{f}_{(i)}(t) = \sum \limits_{k\in\mathbb{Z}} a_{(i)}[k]\varphi(2^{-i}t-k).
\end{equation*}

The expansion coefficients of the orthogonal projection of a signal ${f(t) \in L_2}$ on $\mathcal{V}_{(i)}$ can be determined by filtering and sampling, written in an inner product notation as
\begin{equation}
    \label{eq:OrthogonalProjection}
    a_{(i)}[k] = 2^{-i} \langle f(t), \mathring{\varphi}(2^{-i}t-k) \rangle,
\end{equation}
where $\mathring{\varphi}(t)$ is a dual function of $\varphi(t)$. Since ${\{ \varphi(2^{-i}t-k) \}}$ is a Riesz basis for $L_2$, there is a collection ${\{ \mathring{\varphi}(2^{-i}t-k) \}}$ such that it is biorthogonal to ${\{ \varphi(2^{-i}t-k) \}}$, \textit{i.e.},
\begin{equation*}
    \label{eq:BiorthogonalInnerProduct}
    \langle \varphi(2^{-i}t-k), \mathring{\varphi}(2^{-i}t-l) \rangle = 2^i \delta[k-l].
\end{equation*}
If ${\{ \varphi(2^{-i}t-k) \}}$ is a Riesz basis, then so is its biorthogonal basis \cite{Christensen2003}. The expansion coefficients $a_{(i+1)}[k]$ can be obtained by discrete-time filtering of $a_{(i)}[k]$ and downsampling by a factor of two \cite{Mallat:MRA, Unser:SplinePyramid}. Consequently, all the coefficients $a_{(i)}[k]$ for ${i>0}$ can be computed from $a_{(0)}[k]$ by repeating this procedure.

The loss of information between two successive resolutions ${2^{i}}$ and ${2^{i+1}}$ is called the detail signal at level ${2^{i+1}}$, and is denoted by ${g_{(i+1)} := \hat{f}_{(i)} - \hat{f}_{(i+1)}}$. Let $\mathcal{W}_{(i+1)}$ be the detail subspace at resolution $2^{i+1}$, then ${g_{(i+1)} \in \mathcal{W}_{(i+1)}}$, ${\mathcal{W}_{(i+1)} \perp \mathcal{V}_{(i+1)} }$ and ${\mathcal{W}_{(i+1)} \oplus \mathcal{V}_{(i+1)} = \mathcal{V}_{(i)} }$ \cite{Mallat:MRA}. The subspaces ${\mathcal{W}_{(i+1)}}$ are spanned by vectors $\{ \psi(2^{-i}t - k) \}$, obtained by dilation and translation of a mother wavelet $\psi(t)$. 
The detail signal at resolution $2^{i}$ has the form
\begin{equation*}
    \label{eq:DetailSignalForm}
    g_{(i)}(t) = \sum \limits_{k \in \mathbb{Z}} d_{(i)}[k] \psi(2^{-i}t - k),
\end{equation*}
where $d_{(i)}[k]$ are wavelet coefficients. In order to compute the wavelet coefficients, we can apply the same procedure for obtaining the approximation coefficients $a_{(i)}[k]$, \textit{i.e.},
\begin{equation}
    \label{eq:OrthogonalProjectionWavelets}
    d_{(i)}[k] = 2^{-i} \langle f(t), \mathring{\psi}(2^{-i}t-k) \rangle,
\end{equation}
where $\mathring{\psi}(t)$ is a dual function of $\psi(t)$. The wavelets $\mathring{\psi}(t)$ and $\psi(t)$ are different in the general biorthogonal case, unless the basis functions are orthogonal. For any ${i \in \mathbb{Z}}$, a set of vectors $\{\psi(2^{-(i+1)}t-k)\}$ spans ${\mathcal{W}_{(i+1)} \subset \mathcal{V}_{(i)}}$. Thus, the wavelet coefficients $d_{(i+1)}[k]$ can be computed by discrete-time filtering of $a_{(i)}[k]$ and keeping every other sample of the output \cite{Mallat:MRA}. Finally, for a given signal ${f(t) \in \mathcal{V}_{(0)}}$ in \eqref{eq:SignalinSIspace}, represented by its expansion coefficients $a_{(0)}[k]$ at resolution $2^{0}$, the wavelet decomposition is of the form
\begin{equation}
    \label{eq:WaveletTransform}
    f(t) = \sum \limits_{k \in \mathbb{Z}} a_{(I)}[k] \varphi(2^{-I}t - k) + \sum \limits_{i=1}^{I} \sum \limits_{k\in\mathbb{Z}} d_{(i)}[k]\psi(2^{-i}t - k).
\end{equation}

The choice of scaling and wavelet basis functions is based on certain desirable properties of the multiresolution analysis, whose construction was proposed in numerous papers \cite{Daubechies:Wavelets, Daubechies:OrthonormalWavelets, Vetterli:BiorWavelets}. For image processing, and particularly for coding applications, it is desirable to use symmetric filters with linear phase and compact support. The families of biorthogonal \cite{Cohen:BiOrthogonalWav} and semi-orthogonal \cite{Unser:SplineWavelets} spline wavelets \cite{Unser:TenGoodReasons} satisfy the desired properties, and are efficiently implemented using finite impulse response (FIR) filter banks (see \ref{sec:appendixA}). Their scaling functions are the polynomial B-splines \cite{ Unser:APerfectFit}, whose integer shifts span Riesz bases.

\subsection{Generalized sampling in shift-invariant spaces}
\label{subsec:SamplingInSIspaces}
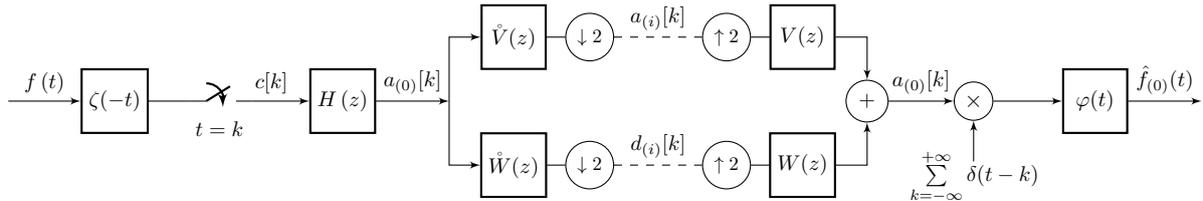
\begin{figure*}[t]
    \centering
    \begin{tikzpicture}[auto, node distance=1cm, >=latex', scale=0.8, every node/.style={scale=0.8}]
        \node [input, name=x_input] (x_input) {};
        \node [block, right of=x_input, node distance=1.75cm] (Prefilter) {$\zeta(-t)$};
        \node [coordinate, right of=Prefilter, node distance=1.25cm](sw_in){};
        \node[coordinate, right of=sw_in](sw_out){};
        \node [coordinate, right of=sw_out] (outputf) {};
        \draw [->] (x_input) -- node{$f\left( t \right)$} (Prefilter);
        \draw [-] (Prefilter) -- (sw_in) ;
        \draw [-] (sw_in) to[cspst] node[anchor=south, xshift=-0.4cm, yshift=-0.7cm]{$t=k$}(sw_out);
        \node [block, right of=sw_out,  node distance=1.5cm] (DigFilt) {$H\left(z\right)$};
        \draw [->] (sw_out) -- node[xshift=-0.15cm]{$c[k]$} (DigFilt);
        \node [coordinate, right of=DigFilt, name=cross, node distance = 1.75cm] (cross) {};
        \draw [->] (DigFilt) -- node{$a_{(0)}[k]$} (cross);
        \node [block, below right of=cross, node distance=1.5cm] (FilterH) {$\mathring{W}(z)$};
        \draw [->] (cross) |- node{}(FilterH);
        \node [sum, right of=FilterH, scale=0.9, node distance = 1.25cm] (downsampleH){$ \downarrow 2$};
        \draw [-] (FilterH) -- node{}(downsampleH);
        \node [input, right of=downsampleH, name=crossH, node distance = 0.75cm] (crossH) {};
        \node [block, above right of=cross, node distance=1.5cm] (FilterL) {$\mathring{V}(z)$};
        \draw [->] (cross) |- node{}(FilterL);
        \node [sum, right of=FilterL, scale=0.9, node distance = 1.25cm] (downsampleL){$\downarrow 2$};
        \draw [-] (FilterL) -- node{}(downsampleL);
        \node [input, right of=downsampleL, name=crossL, node distance = 0.75cm] (crossL) {};
        \node [sum, right of=downsampleH, scale=0.9, node distance = 2.25cm] (upsampleHrec){$\uparrow 2$};
        \node [block, right of=upsampleHrec, node distance=1.25cm] (FilterHrec) {$W(z)$};
        \draw [-] (upsampleHrec) -- node{}(FilterHrec);
         \node [sum, right of=downsampleL, scale=0.9, node distance = 2.25cm] (upsampleLrec){$\uparrow 2$};
        \node [block, right of=upsampleLrec, node distance=1.25cm] (FilterLrec) {$V(z)$};
        \draw [-] (upsampleLrec) -- node{}(FilterLrec);
        \draw [-, dashed] (downsampleL) -- node{$a_{(i)}[k]$}(upsampleLrec);
        \draw [-, dashed] (downsampleH) -- node{$d_{(i)}[k]$}(upsampleHrec);
        \node [sum, above right of=FilterHrec, name=crossOut, node distance = 1.5cm] (crossOut) {$\suma$};
        \draw[->] (FilterHrec) -| node{}(crossOut);
        \draw[->] (FilterLrec) -| node{}(crossOut);
        \node [sum, right of=crossOut, scale=0.75, node distance = 1.75cm] (mult) {\mult};
        \draw[->] (crossOut) -- node{$a_{(0)}[k]$}(mult);
        \node [input, name=delta_input, below of=mult, node distance=1cm] (delta_input) {2};
        \draw [->] (delta_input) node[anchor=east,xshift=1.2cm, yshift=-0.25cm]{$\sum \limits_{k=-\infty}^{+\infty}     \delta(t-k)$}-- (mult);
        \node [block, right of=mult,node distance=2cm] (RecFilt){$\varphi(t)$};
        \draw [->] (mult) -- (RecFilt);
        \node [output, right of=RecFilt, node distance=1.75cm] (output) {};
        \draw [->] (RecFilt) -- node {$\hat{f}_{(0)}(t)$}(output);
    \end{tikzpicture}
    \caption{Sampling and reconstruction of a signal that lies in a shift-invariant subspace with a perfect reconstruction wavelet filter bank for multiresolution analysis and compression. The signal is filtered with an arbitrary sampling kernel $\zeta(t)$ before being sampled and filtered with a discrete-time correction filter $H(z)$ to obtain the expansion coefficients $a_{(0)}[k]$. Notice that the role of $\zeta(t)$ is analogous to that of an antialiasing filter in the Shannon's sampling framework.}
    \vspace{-1.75ex}
    \label{fig:SamplingInSIspaces}
\end{figure*}

Before applying the exact wavelet decomposition by using the discrete-time analysis filters, the signal should initially be represented by appropriate approximation coefficients at resolution $2^{0}$. We assume that a continuous-time signal $f(t)$ lies in an SI subspace $\mathcal{V}_{(0)}$ of $L_2$ and is given by \eqref{eq:SignalinSIspace}. The generator $\varphi(t)$ forms a Riesz basis for $L_2$ and satisfies \eqref{eq:RieszBasisCondition}. In order to obtain a discrete-time representation, the signal is filtered with an analog prefilter and sampled uniformly at time instants $k$. Samples $c[k]$ in the general SI sampling scheme (see Figure \ref{fig:SamplingInSIspaces})
can be expressed as
\begin{equation}
    \label{eq:SamplingInSIdef}
    c[k] = \int \limits_{-\infty}^{\infty}f(t)\zeta(t-k)dt \triangleq \langle f(t), \zeta(t-k) \rangle,
\end{equation}
where $\zeta(-t)$ is the impulse response of the prefilter. Sampling in SI spaces \cite{Unser:GeneralSampTheory, Unser:50YearsAfterShannon, Eldar:BeyondBandlimitedSampling} is practical and retains the shift-invariant property of the Shannon's sampling theory. In an ideal scenario, the sampling kernel $\zeta(t)$ is orthogonal or biorthogonal to the signal generator $\varphi(t)$, which implies that ${c[k]=a_{(0)}[k]}$. However, in practical applications sampling is often imposed by a physical device. Thus, a more realistic setting is to let $\zeta(t)$ be an arbitrary kernel.

Let us denote the sampled cross-correlation sequence between the sampling kernel $\zeta(t)$ and the signal generator $\varphi(t)$ with
\begin{equation}
    \label{eq:CrossCorrelationSeq}
    r[k]=\langle \varphi(t), \zeta(t-k) \rangle. 
\end{equation}
The expression \eqref{eq:SamplingInSIdef} of the samples $c[k]$ is equal to
\begin{equation*}
    \label{eq:SamplesExpandedShort}
    c[k] = a_{(0)} * r[k],
\end{equation*}
where $(*)$ is the convolution operator (see \ref{subsec:appendixB1} for the derivation). Using the Fourier relations, it follows that ${C(e^{j\omega}) = A_{(0)}(e^{j\omega}) R(e^{j\omega})}$, where $C(e^{j\omega})$, $A_{(0)}(e^{j\omega})$ and $R(e^{j\omega})$ are the discrete-time Fourier transforms (DTFTs) of $c[k]$, $a_{(0)}[k]$ and $r[k]$, respectively. In order to obtain $a_{(0)}[k]$, the orthogonal projection of $f(t)$ onto $\mathcal{V}_{(0)}$ can be obtained from samples $c[k]$ by filtering with a correction filter whose impulse response is given by the DTFT  \cite{Unser:GeneralSampTheory, Unser:50YearsAfterShannon, Eldar:BeyondBandlimitedSampling} \begin{equation}
    \label{eq:CorrFilter}
    H(e^{j\omega}) = \frac{1}{R(e^{j\omega})}.
\end{equation}
A rather mild condition on the generator $\varphi(t)$ and the sampling kernel $\zeta(t)$ should be met in order to be able to recover $a_{(0)}[k]$ from the samples \cite{Eldar:BeyondBandlimitedSampling}, \textit{i.e.}, ${\left\vert R(e^{j\omega}) \right\vert>0}$. Note that for orthogonal and biorthogonal basis functions ${r[k]=\delta[k]}$ and ${H(e^{j\omega}) = 1}$. When orthogonality or biorthogonality is not satisfied, an elegant SI sampling framework and an efficient implementation of the correction filter \eqref{eq:CorrFilter} is possible in B-spline function spaces \cite{Unser:50YearsAfterShannon, Unser:BsplineSigProc, Unser:BsplineSigProcII}. In such function spaces, the correction filter can be determined analytically and the denominator in \eqref{eq:CorrFilter} corresponds to a concatenation of a causal and an anticausal infinite impulse response (IIR) filter with a simple expression \cite{Unser:BsplineSigProc}. This is due to the compact support of the polynomial B-splines for which the cross-correlation sequence $r[k]$ has only a few nonzero symmetric entries around ${k=0}$. 

Once the approximation coefficients $a_{(0)}[k]$ are determined, the discrete wavelet transform (DWT) described in Section \ref{subsec:MultiresRepresentation} can be applied by using a perfect reconstruction filter bank. Notice that $a_{(0)}[k]$ are not necessarily pointwise values of the signal. To reconstruct the signal, $a_{(0)}[k]$ are modulated with a train of Diracs $\sum_{k\in\mathbb{Z}}\delta(t-k)$ and filtered with an analog filter whose impulse response corresponds to the generator $\varphi(t)$. The reconstruction $\hat{f}_{(0)}(t)$ is the minimum squared error approximation of the signal $f(t)$ in $\mathcal{V}_{(0)}$. The reconstruction procedure is illustrated in Figure \ref{fig:SamplingInSIspaces}. However, it is often more desirable to reconstruct the pointwise values of the signal in various resolutions than its analog representation, especially for two-dimensional signals. The pointwise values of $\hat{f}_{(0)}(t)$ is particularly efficient to obtain by using a B-spline representation \cite{Unser:BsplineSigProc, Unser:BsplineSigProcII}. In this case, instead of modulation with a train of Diracs and analog filtering, the recovered approximation coefficients $a_{(0)}[k]$ are zero-stuffed and filtered with an FIR filter that has only a few coefficients.

\subsection{Extensions to higher dimensions}
\label{subsec:Extensions2HiDim}
The results in Section \ref{subsec:MultiresRepresentation} and Section \ref{subsec:SamplingInSIspaces} can be extended to higher dimensions through the use of the tensor product, provided that the sampling is performed on the Cartesian grid \cite{Mallat:MRA, Unser:50YearsAfterShannon, Unser:BsplineSigProc}. The formulas can be extended by considering the space variables ${(u,v) \in \mathbb{R}^2}$, and by replacing single summations and integrals with multiple ones. In practice, we often use separable functions, which greatly simplify the implementation. Separability retains the one-dimensional Riesz basis condition in multiple dimensions \cite{Unser:50YearsAfterShannon}. 

A signal ${f(u,v) \in \mathcal{V}_{(0)}^2}$ has the form
\begin{equation}
    \label{eq:2DsignalSIspace}
    f(u,v) = \sum\limits_{k\in\mathbb{Z}} \sum\limits_{l\in\mathbb{Z}} a_{(0)}[k,l] \varphi(u-k)\varphi(v-l),
\end{equation}
where $a_{(0)}[k,l]$ are two-dimensional expansion coefficients. Furthermore, two-dimensional wavelet decomposition is obtained by successive one-dimensional fast wavelet transform along the rows and columns \cite{Mallat:MRA, Unser:SplineWavelets}.

The two-dimensional samples $c[k,l]$ can be expressed as
\begin{equation}
    \label{eq:SIsamplesIn2D}
    c[k,l] = (a_{(0)} \; {**} \; r_k) \; {**} \; r_l[k,l],
\end{equation}
where ${(**)}$ is a two-dimensional convolution operator. The convolution is separable, \textit{i.e.}, $r_k$ and $r_l$ denote row and column vectors composed of the cross-correlation sequence $r[k]$. A detailed explanation of the result in \eqref{eq:SIsamplesIn2D} is given in \ref{subsec:appendixB2}. It follows that all filtering operations are separable and applied successively along the coordinates.

\section{Single-pixel imaging via compressive sensing in shift-invariant spaces}
\label{sec:CSinSIspaces}
Sampling and reconstruction in SI spaces resembles the Nyquist-Shannon theorem. The sampling rate in the SI sampling framework is determined by a number of degrees of freedom per time unit, which is also referred to as the \textit{rate of innovation} \cite{Vetterli:FRI}. Signals are sampled at least at the rate of innovation ${\rho=1/T=1}$, for ${T=1}$, in order to represent $f(t)$ exactly. Signals in SI spaces are often related to high rates of innovation, even though their \textit{information rates} (\textit{e.g.}, the number of nonzero coefficients in a certain transform domain) might be low. It is challenging to build sampling hardware that operates at a sufficient rate when the rate of innovation is high. For example, high-resolution imaging technologies in the non-visible wavelength spectrum can be impractical or very expensive to implement.

Compressive sensing \cite{Candes:CS, Donoho:CS, Foucart:IntroCS} is a sampling and reconstruction framework that, unlike the conventional sampling theorems, focuses on the information rate of the signal rather than its innovation rate. A signal with a low information rate can be exactly recovered from far fewer measurements (samples) in the CS framework than in the conventional sampling frameworks, even if it has a high innovation rate. The theory of CS has led to the enhancement of signal reconstructions in low-resolution imaging devices \cite{Ralasic:DualImaging} and in medical imaging modalities where the full measurements are missing or not available (\textit{e.g.}, magnetic resonance imaging \cite{Lustig:MRI}, computed tomography, etc.).

The majority of papers concerning CS rely on a discretization of the measurement process or heuristics to adopt a continuous-domain inverse problem to a finite-dimensional CS setting. A discrete measurement model of a continuous measurement process is often an approximation that can lead to bad signal reconstructions. Furthermore, it is typically assumed that the observed signal lies in a pixel-by-pixel basis and can be represented by its pointwise values. If not specifically taken care of in the discretization procedure, such an assumption additionally contributes to the error related to the reconstruction of the underlying continuous-domain signal. Furthermore, the Nyquist-Shannon model of the underlying signal may not be the best solution, since signals are often much better represented in other SI bases, such as splines \cite{Unser:50YearsAfterShannon, Eldar:BeyondBandlimitedSampling}. Recently, a framework for CS of one-dimensional signals that lie in an arbitrary SI subspace was reported in \cite{Vlasic:EUSIPCO} and \cite{Vlasic:CSinSIspaces}. The framework links the theories of generalized sampling in SI spaces and CS, leading to an exact discretization of continuous-domain inverse problems in devices specialized for CS where the sampling kernel spans an SI subspace (\textit{e.g.}, the random demodulator \cite{Tropp:BeyondNyquist}).

\subsection{Single-pixel compressive imaging in SI spaces}
\label{subsec:SinglePxCIinSIspaces}
We reinterpret the single-pixel camera as a system for acquisition of two-dimensional signals given by \eqref{eq:2DsignalSIspace} that lie in an arbitrary SI subspace. The SI function subspaces are used to model the underlying continuous-domain measurement patterns and input signals, and the principles of generalized sampling are used to discretize the continuous measurement procedure. The discretization method is exact and avoids approximations that lead to reconstruction errors in standard single-pixel imaging.

\begin{figure}[t]
    \centering
    \includegraphics[width=0.975\columnwidth]{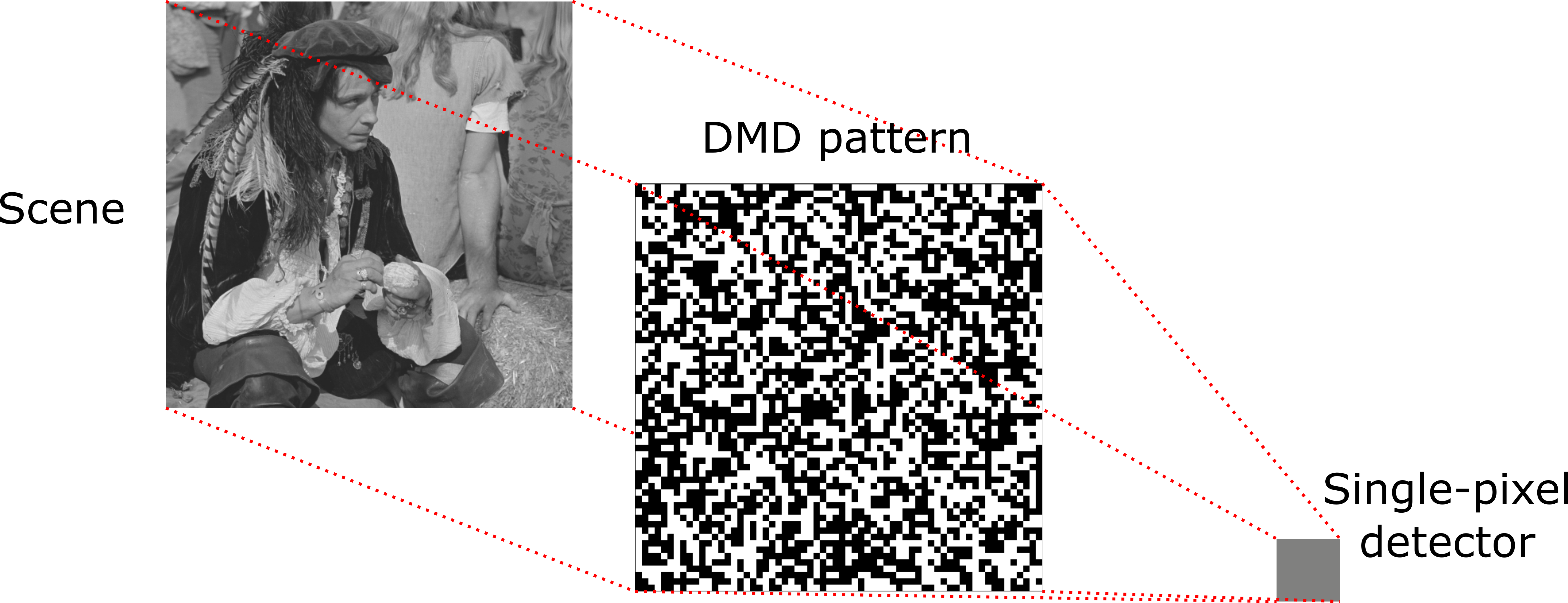}
    \caption{\textbf{Single-Pixel Imaging.} A scene is modulated by a measurement pattern and the single-pixel detector measures the intensity of light beams that reaches its surface.}
    \vspace{-1.5ex}
    \label{fig:SinglePxCamera}
\end{figure}

Single-pixel imaging modulates a scene with a sequence of spatially resolved patterns and measures the intensities of the correlations between the patterns and the scene \cite{Duarte:SinglePxCamera, Edgar2019, Gibson2020}. A simple illustration of single-pixel imaging is given in Figure \ref{fig:SinglePxCamera}.
Typically, the modulation is implemented by the DMDs, consisting of an array of individually addressable micromirrors that redirects the illumination or the light beams. The DMD is implemented as a programmable binary mask, where the value of one corresponds to full illumination of a scene (or redirection of a beam to the detector), and the value of zero means that the illumination (or the light beam) is blocked. A common approach is to use a sequence of orthogonal binary masks, \textit{e.g.}, the Hadamard basis, or $\pm1$ random Bernoulli masks, and measure the differential intensity for each pattern and its photographic negative. The single-pixel detector integrates the light photons which fall on its surface and converts them into electrons whose number will be directly proportional to the intensity of the modulated scene.

Let us define the box function, which is also referred to as the B-spline of order zero, as:
\begin{equation*}
    \label{eq:boxFunc}
    b^0(t) =    \begin{cases}
			         1,	& 0 < t \leq 1 \\
			         0, & \mathrm{otherwise}
		        \end{cases}.
\end{equation*}
Its integer shifts satisfy condition \eqref{eq:RieszBasisCondition}. The DMD measurement masks can be modeled as functions $\{z_m(u,v)\}^{M}_{m = 1}$ that lie in an SI subspace $\mathcal{Z}$ spanned by the integer shifts of a separable sampling kernel ${\zeta(u,v) = b^0(u)b^0(v)}$. Any spatially resolved pattern in $\mathcal{Z}$ is given in the form
\begin{equation*}
    \label{eq:MeasurementFunction}
    z_m(u,v) = \sum_{k=0}^{K-1} \sum_{l=0}^{L-1} s_{m}[k,l] \zeta(u-k, v-l),
\end{equation*}
where $s_{m}[k,l]$ are the expansion coefficients of values ${\pm 1}$, and the ${(K,L)}$ pair determines the resolution of the measurement masks. By assuming that a scene model ${f(u,v) \in \mathcal{V}_{(0)}^2}$ is given by \eqref{eq:2DsignalSIspace}, a single measurement $y_m$ of a detector in the single-pixel imaging system is of the form
\begin{align*}
        y_m &= \int \limits_{u=0}^{K} \int \limits_{v=0}^{L} f(u,v)z_m(u,v) dudv \\
            &= \int \limits_{u=0}^{K} \int \limits_{v=0}^{L} \left (\sum\limits_{i\in\mathbb{Z}} \sum\limits_{j\in\mathbb{Z}} a_{(0)}[i,j] \varphi(u-i)\varphi(v-j) \right. \nonumber \\
            & \qquad \qquad \quad \ \times \sum_{k=0}^{K-1} \sum_{l=0}^{L-1} s_{m}[k,l] \zeta(u-k) \zeta(v-l) \Bigg)dudv. \nonumber
\end{align*}
The integration interval ${[0,K] \times [0,L]}$ covers the whole support of measurement masks, \textit{i.e.}, it covers the support of exactly ${K \times L}$ basis functions that span the space $\mathcal{Z}$. Thus, we can use the inner product notation and the results given in \ref{subsec:appendixB2} to express the measurement by
\begin{align*}
    y_m &= \sum_{k=0}^{K-1} \sum_{l=0}^{L-1} s_{m}[k,l] \sum_{i\in \mathbb{Z}} \sum_{j\in \mathbb{Z}} a_{(0)}[i,j] r[k-i] r[l-j] \nonumber \\
        &= \sum_{k=0}^{K-1} \sum_{l=0}^{L-1} s_{m}[k,l] (a_{(0)} \; {**} \; r_k) \; {**} \; r_l[k,l]
\end{align*}
where $r[k]$ is the sampled cross-correlation sequence between the one-dimensional signal generator $\varphi(t)$ and the one-dimensional sampling kernel ${\zeta(t)=b^0(t)}$, defined in \eqref{eq:CrossCorrelationSeq}.
Finally, by applying \eqref{eq:SIsamplesIn2D}, the measurement is given by a weighted linear combination of the samples in generalized SI sampling framework for B-spline of order zero as the sampling kernel:
\begin{equation*}
    \label{eq:MeasurementsLinComb}
    y_m = \sum_{k=0}^{K-1} \sum_{l=0}^{L-1} s_{m}[k,l] c[k,l].
\end{equation*}
Measurements ${\bm{y} \in \mathbb{R}^M}$ are given in a matrix form as:
\begin{equation}
    \label{eq:MeasurementsInMtxForm}
    \bm{y} = \bm{S c},
\end{equation}
where $\bm{S}$ is an ${M \times N}$ \textit{measurement matrix}, for ${N=KL}$, whose rows consist of the coefficients $\{s_m[k,l]\}_{m=1}^{M}$. Consequently, ${\bm{c} \in \mathbb{R}^{N}}$ is a vector form of the samples $c[k,l]$. In single-pixel imaging, the number of measurement patterns is often ${M<N}$, leading to an ill-posed system of equations. The theory of CS has been extensively used in order to reconstruct the image from a reduced set of measurements. In the next section, we will offer a solution for image reconstruction from $\bm{y}$ by using CS via exact wavelet frames that intuitively complements the proposed model of the underlying signal.

In the conventional single-pixel imaging setting, the $\bm{c}$ samples in \eqref{eq:MeasurementsInMtxForm} are considered as pointwise values $f(k,l)$ of the signal, which is based on the assumption that the underlying signal is bandlimited, \textit{i.e.}, ${\varphi(u,v)=\mathrm{sinc}(u)\mathrm{sinc}(v)}$. By considering the measurement model proposed in this paper, such a discretization would be just an approximation since ${\varphi(t)=\mathrm{sinc}(t) \not \perp \zeta(t)=b^0(t)}$, and thus ${r[k] \neq \delta[k]}$ and ${c[k,l] \neq a_0[k,l] = f(k,l)}$. However, the conventional single-pixel imaging problem is exact if $\varphi(t)$ is assumed to be the box function, \textit{i.e.}, if ${\varphi(t)=\zeta(t)=b^0(t)}$, since it is orthonormal to its integer shifts (${r[k]=\delta [k]}$) and since its expansion coefficients $a_0[k,l]$ correspond to the pointwise values $f(k,l)$ of the signal. The proposed framework allows one to use an SI signal model and generalized sampling theory which suit better to the observed underlying signal and lead to exact discretization and reconstruction.

\subsection{Wavelet-domain sparse reconstruction}
\label{subsec:SparseWavRec}
Measurements $\bm{y}$ in \eqref{eq:MeasurementsInMtxForm} can be expressed in the form
\begin{equation}
    \label{eq:MeasConvMtx}
    \bm{y} = \bm{S R \, a}_{(0)},
\end{equation}
where $\bm{R}$ is a convolution matrix representing an implementation of the two-dimensional convolution with the sampled cross-correlation sequence $r[k]$, and $\bm{a}_{(0)}$ is a vector form of the expansion coefficients $a_{(0)}[k,l]$. If we denote the number of nonzero elements in $r[k]$ with ${\Omega = \mathrm{card} \{r[k] \; | \; r[k] \neq 0 \}}$, where $\mathrm{card}$ is a set cardinality, the dimension of the matrix $\bm{R}$ is ${N \times \tilde{N}}$, for ${\tilde{N} = (K+\Omega-1)(L+\Omega-1)}$, and consequently, $\bm{a}_{(0)}$ is an ${\tilde{N} \times 1}$ vector.

Numerous papers \cite{Eldar:CSAnalogSI, Eldar:RobustRecovery, Unser:SplinesUnSolution, Debarre2019} assume that the observed signal is sparse in a certain SI basis. That is, these works use a prior that the signal can be represented with only a few nonzero expansion coefficients in the SI basis, such as a B-spline function space. We can exploit the same assumption in combination with our measurement model in \eqref{eq:MeasConvMtx} and recover the coefficients $\bm{a}_{(0)}$ by solving \eqref{eq:Intro_CsProblem}, for ${\bm{\Theta}:=\bm{SR}}$ and ${\bm{x}:=\bm{a}_{(0)}}$. However, we take a different approach.

In this paper, an SI basis $\varphi(u,v)$ provides an underlying model of the observed continuous-domain signal that may not be sparse when represented with the corresponding approximation coefficients $a_{(0)}[k,l]$. Instead, $a_{(0)}[k,l]$ are assumed to be sparse in a certain sparsity basis $\bm{\Psi}$. While the choice of the basis can be arbitrary, and usually based on a signal prior \cite{Vlasic:CSinSIspaces}, an intuitive option is a wavelet basis that complements the scaling function $\varphi(u,v)$. Moreover, such a pair of wavelet and scaling functions leads to an exact discretization of a continuous-domain inverse problem into a finite CS setting, where the recovered sequence corresponds to the coefficients obtained by the wavelet decomposition. While CS reconstruction with a wavelet-based sparsity domain has been considered previously \cite{LihanHe:WBBCS, Baraniuk:ModelBasedCS, Torkamani2021}, the resulting methods assume the bandlimited signal model whose discrete pointwise values ($\bm{\Psi x}$) are reconstructed by solving \eqref{eq:Intro_CsProblem}. Here, we use an SI signal model whose generator is a scaling function corresponding to a specific wavelet, which leads to the reconstruction of the approximation coefficients $a_{(0)}[k,l]$ of the continuous-domain signal. 

Consider a signal $f(u,v) \in \mathcal{V}_{(0)}^2$ given by \eqref{eq:2DsignalSIspace}, whose wavelet decomposition is a two-dimensional version of \eqref{eq:WaveletTransform}. By using discrete-domain synthesis filters $v[k]$ and $w[k]$ whose $z$-transforms are $V(z)$ and $W(z)$ (see Figure \ref{fig:WaveletFilterBank}), one can construct a synthesis inverse discrete wavelet transform (IDWT) matrix ${\bm{\Psi} \in \mathbb{R}^{\tilde{N}\times\tilde{N}}}$. In the paper, we use a traditional dyadic wavelet reconstruction model. For polynomial B-spline scaling functions, spline wavelets \cite{Cohen:BiOrthogonalWav, Unser:TenGoodReasons} are sparsity inducing complements that can efficiently be implemented in an exact transform matrix due to the corresponding FIR filters $V(z)$ and $W(z)$. The approximation coefficients $\bm{a}_{(0)}$ of the signal $f(u,v)$ are then given by
\begin{equation}
    \label{eq:ApproxCoeffsAndSparsityMtx}
    \bm{a}_{(0)} = \bm{\Psi x},
\end{equation}
where ${\bm{x} = [\bm{a}_{(I)}^T, \bm{d}_{u(I)}^T, \bm{d}_{v(I)}^T, \bm{d}_{uv(I)}^T, \bm{d}_{u(I-1)}^T, \dots , \bm{d}_{uv(1)}^T]^T}$ is an ${\tilde{N} \times 1}$ vector of two-dimensional wavelet decomposition coefficients. In such a setting, the coefficients in $\bm{x}$ correspond exactly to the approximation and wavelet coefficients in two-dimensional versions of \eqref{eq:OrthogonalProjection} and \eqref{eq:OrthogonalProjectionWavelets}. The measurements $\bm{y}$ in \eqref{eq:MeasConvMtx} can be expanded to
\begin{equation*}
    \bm{y} = \bm{S R \, \Psi x},
\end{equation*}
where ${\bm{\Theta}:=\bm{SR\,\Psi}}$ is an ${M \times \tilde{N}}$ sensing matrix. In conventional CS, measurements $\bm{y}$ are usually approximated with ${\bm{S \, \Psi x}}$ which, by considering the proposed measurement setup, leads to recovery of SI samples ${\bm{c}=\bm{\Psi x}}$ (see \eqref{eq:MeasurementsInMtxForm}). The recovered vector $\bm{c}$ is then considered as a final solution which is an error unless ${\langle \varphi(t), \zeta(t-k) \rangle = \delta[k]}$, implying that the observed continuous-domain signal lies in a piecewise constant space spanned by the integer shifts of the box function $b^0(u,v)$. In our case, the $\bm{R}$ matrix allows for direct recovery of approximation coefficients $\bm{a}_{(0)}$ that represent the minimum squared error approximation of the observed signal in $\mathcal{V}_{(0)}^2$. This offers an exact discretization procedure of the inherently continuous inverse problem with opportunity to model the underlying signal with an appropriate generator $\varphi(u,v)$.

We exploit the sparsity prior by using CS and reconstruct the coefficients by solving an optimization problem
\begin{equation}
    \label{eq:CSrecSIspaces}
    \arg \min_{\bm{x}\in \mathbb{R}^{\tilde{N}}} \Vert \bm{y} - \bm{S R \, \Psi x} \Vert_{\ell_2}^2 + \lambda \Vert \bm{x} \Vert_{\ell_1},
\end{equation}
where the $\ell_1$-norm regularization promotes sparse solutions \cite{Donoho:L1Sparsest} and $\lambda$ is a regularization parameter. Traditional CS is concerned with the RIP, a strict condition on a sensing matrix $\bm{\Theta}$ that preserves the geometry and guarantees an exact recovery of the $Q$-sparse vector $\bm{x}$ \cite{Candes:CS, Candes:RIP}. However, over the years, the RIP has proven to be too strong an assumption, especially in practical CS where the exact recovery is almost impossible to achieve \cite{Bastounis:AbsenceOfRIP, Adcock:BreakingCoherence}. Instead, a rather mild condition, similar to the Riesz-basis condition, is enough to be met to ensure a stable recovery \cite{Eldar:SampTheory}
\begin{equation}
    \label{eq:CSmildRIP}
    \alpha \left\Vert \bm{x} \right\Vert_{\ell_2}^{2} \leq \left\Vert \bm{\Theta x} \right\Vert_{\ell_2}^{2} \leq \beta \left\Vert \bm{x} \right\Vert_{\ell_2}^{2},
\end{equation}
for constants ${0<\alpha \leq \beta<\infty}$. In this paper, the sampling kernel and the generator are imposed to form Riesz bases, and thus ${\bm{\Theta} := \bm{S R \, \Psi}}$ satisfies the condition in \eqref{eq:CSmildRIP}. A property of $\bm{\Theta}$ that provides a concrete measure of its recovery ability is the coherence between the measurement and sparsity matrices. The coherence is defined as the maximal correlation between two matrices $\mathbf{G}$ and $\mathbf{H}$ as
\begin{equation*}
    \label{eq:Coherence}
    \mu(\mathbf{G},\mathbf{H}) \triangleq \max_{\substack{1 \leq i \leq M \\ 1 \leq j \leq \tilde{N}}} \frac{\vert g_i h_j \vert}{ \Vert g_i \Vert_{\ell_2} \Vert h_j \Vert_{\ell_2} },
\end{equation*}
where $\{g_i\}$ and $\{h_j\}$ are rows and columns of $\mathbf{G}$ and $\mathbf{H}$, respectively. Small values of the coherence are preferable as they allow reconstruction of denser vectors $\bm{x}$ for a fixed number of measurements $M$. Random matrices are largely incoherent with any deterministic sparsity matrix \cite{Candes:CS, Canh2021}. Thus, random Gaussian and Bernoulli matrices provide universal solutions to the problem of choosing an appropriate measurement matrix in the CS setting. The implementation of measurement masks with random i.i.d. values of $\pm 1$ in single-pixel imaging is straightforward with the DMD. In the proposed framework, the sparsity matrix is $\bm{\Psi}$, and $\bm{S}$ and $\bm{R}$ together form a measurement matrix, since $\bm{R}$ is based on the measurement setup and the underlying signal model. The measurement matrix depends on the cross-correlation between the sampling kernel and the signal generator, which introduces a deterministic structure within the random entries of $\bm{S}$. This may increase the coherence of the sensing matrix ${\bm{\Theta}:=\bm{SR \, \Psi}}$ in comparison with the coherence between the conventional pair ${\mu(\bm{S},\bm{\Psi})}$. However, for specific SI basis functions used in the experiments, random measurement matrices are adequate choice that secures low coherence and quality reconstruction results from a small number of CS measurements.

The proposed framework allows for sampling of signals in SI subspaces with a much lower sampling rate in contrast to the conventional high-rate method described in Section \ref{subsec:SamplingInSIspaces} and Figure \ref{fig:SamplingInSIspaces}. Single-pixel imaging and recovery via solving the optimization problem \eqref{eq:CSrecSIspaces} replaces high-resolution imaging and compression procedure, and directly recovers wavelet coefficients, \textit{i.e.}, expansion coefficients $\bm{a}_{(0)}$, that are further to be stored or processed. Notice that the recovered coefficients $\bm{a}_{(0)}$ are not necessarily the pointwise values of the signal reconstruction. They can be seen as a parametric representation of the continuous-domain signal obtained directly from a reduced set of measurements. When represented in an appropriate SI basis, the recovered signal can be processed directly in that domain without the need of transforming the coefficients into the pointwise values. B-splines are an alternative approach to the bandlimited theory which offers many practical advantages in image processing \cite{Unser:APerfectFit}. The underlying signal model are polynomials that are more suitable for operations like convolution or differentiation than computing discrete approximations to these mathematical constructions. Such operations are efficiently implemented with FIR filters \cite{Unser:BsplineSigProc, Unser:BsplineSigProcII}. If one desires a pointwise representation of the signal, it can also be obtained by FIR filtering of the recovered coefficients $\bm{a}_{(0)}$, as described in Section \ref{subsec:SamplingInSIspaces}. Furthermore, B-splines are scaling functions of the biorthogonal wavelets \cite{Cohen:BiOrthogonalWav} that are commonly used in the coding applications (Section \ref{subsec:MultiresRepresentation}). The biorthogonal wavelets are an intuitive complement of the B-spline basis functions and will be used to construct sparsity domains in the rest of the paper. However, notice that the proposed framework is not solely limited to the biorthogonal wavelets nor B-spline basis functions.

\section{Algorithmic details and implementation}
\label{sec:Implementation}
It is quite costly to implement random measurement matrices in practical large-scale compressive imaging applications since they require huge memory allocation, because of their unstructured nature, and necessitate many multiplications in the order of $\mathcal{O}(MN)$. There are several articles proposing methods to overcome these issues and, in this paper, we realize measurement matrices by implementing the structurally-random-matrix (SRM) framework \cite{Do:SRM}. We use the matrix-free interior point method (mfIPM) \cite{Fountoulakis:MfIPM} for solving large-scale CS problems, which exploits the ability of the signal processing matrices to perform inexpensive matrix-vector multiplications that are typically in the order of ${\mathcal{O}(\tilde{N})}$ to ${\mathcal{O}(\tilde{N}\log \tilde{N})}$. Finally, the proposed CS framework is conducted on the standard test images and measurements obtained by the system proposed in \cite{Ralasic:OffTheShelf}.

\subsection{Structurally random measurement matrix}
\label{subsec:StructRandMtx}
A structurally random measurement matrix $\bm{S}$ is defined as a product of three matrices \cite{Do:SRM}:
\begin{equation*}
    \label{eq:SRMM}
    \bm{S} = \bm{D F P},
\end{equation*}
where
\begin{itemize}[noitemsep, nolistsep]
  \item ${\bm{P} \in \mathbb{R}^{N \times N}}$ is a uniform random permutation matrix which scrambles the signal's sample locations.
  \item ${\bm{F} \in \mathbb{R}^{N \times N}}$ is a Walsh-Hadamard transform (WHT) matrix, in that SRM's entries only take values of $\pm1$ for compatibility with the hardware implementation.
  \item ${\bm{D} \in \mathbb{R}^{M \times N}}$ is a subsampling matrix which randomly picks up $M$ rows of the matrix $\bm{FP}$.
\end{itemize}
These matrices can be efficiently realized with their non-matrix counterparts that require only a small amount of memory storage. The costs of permutation and random subsampling are in the orders of $\mathcal{O}(N)$ and $\mathcal{O}(M)$, respectively. The fast WHT is efficiently realized with complexity in the order of ${\mathcal{O}(N\log N)}$ by only using the operations of additions and subtractions. The SRM performance in the CS setting is theoretically on average comparable to that of the completely random matrix, leading to a low-coherent pair of measurement and sparsity matrices \cite{Do:SRM}.

\subsection{Matrix-free reconstruction algorithm}
\label{subsec:MtxFreeRecMethod}
In this paper, we use mfIPM to recover the observed signal from a reduced set of measurements. The solver is robust and designed for signal reconstruction problems that arise in the filed of CS \cite{Fountoulakis:MfIPM}. The mfIPM, as well as many other iterative $\ell_1$-minimization problem solvers, spends the majority of computational time to determine matrix-vector multiplications ${\bm{\Theta x}}$ and ${\bm{\Theta}^T\bm{x}}$, where $\bm{\Theta}$ is a sensing matrix. In the proposed framework, by using the notation of SRM, the problem of computing $\bm{\Theta x}$ is given by
\begin{equation}
    \label{eq:ForwardModel}
    \bm{\Theta x} = \bm{D F P R \, \Psi x},
\end{equation}
where all of the matrices have their non-matrix counterparts. Recall that $\bm{R}$ is a matrix form of the separable two-dimensional convolution with the cross-correlation sequence $r[k]$, and $\bm{\Psi}$ is a matrix form of the two-dimensional synthesis IDWT. The costs of the two-dimensional separable convolution and fast IDWT are in the order of $\mathcal{O}(\Omega \tilde{N})$ and  $\mathcal{O}(\tilde{N})$, respectively. Notice that the number of nonzero elements $\Omega$ in the sequence $r[k]$ is rather small when the signal generator is a B-spline basis function of a low order $p$. The entries of the cross-correlation sequence $r[k]$ for B-spline signal generators of orders ${p=0,\dots,3}$ are given in Table \ref{tab:CrossCorrelationSeq}.
\begin{table}[b]
    \vspace{-1.65ex}
	\caption{Sequence $r[k]$ for B-Spline Signal Generator $\varphi(t)$ of Order $p$}
	\label{tab:CrossCorrelationSeq}
	\centering
	\begin{tabular}{c | c  | c}
			\multicolumn{1}{c |}{$p$} & \multicolumn{1}{c |}{$\varphi(t)$} & \multicolumn{1}{c}{$r[k]$} \\
			\hline
			\multicolumn{1}{c |}{} & \\[-1em]
			\multicolumn{1}{c |}{$0$} & \multicolumn{1}{c |}{$b^0(t)$} & \multicolumn{1}{c}{$\{ \dots,0,\underline{1},0,\dots \}$}\\
			\multicolumn{1}{c |}{} & \\[-1em]
			\multicolumn{1}{c |}{$1$} & \multicolumn{1}{c |}{$b^1(t)$} & \multicolumn{1}{c}{$\{ \dots,0,\frac{1}{8},\underline{\frac{3}{4}},\frac{1}{8},0,\dots \}$}\\
			\multicolumn{1}{c |}{} & \\[-1em]
			\multicolumn{1}{c |}{$2$} & \multicolumn{1}{c |}{$b^2(t)$} & \multicolumn{1}{c}{$\{ \dots,0,\frac{1}{6},\underline{\frac{2}{3}},\frac{1}{6},0,\dots \}$}\\
			\multicolumn{1}{c |}{} & \\[-1em]
			\multicolumn{1}{c |}{$3$} & \multicolumn{1}{c |}{$b^3(t)$} & \multicolumn{1}{c}{$\{ \dots,0,\frac{1}{384},\frac{76}{384},\underline{\frac{230}{384}},\frac{76}{384},\frac{1}{384},0,\dots \}$}
	\end{tabular}
	\vspace{-1.65ex}
\end{table}
For the B-spline signal generator of order $0$, the cross-correlation sequence ${r[k] = \delta[k]}$, since the sampling kernel and the generator are orthonormal. This results in ${\bm{R} = \bm{I}}$, where $\bm{I}$ is an identity matrix, and the signal reconstruction corresponds to the conventionally discretized problem of CS type where ${\bm{\Theta}:=\bm{S \Psi}}$.
Fast IDWT is efficiently computed by FIR filtering for synthesis spline wavelets of compact support \cite{Cohen:BiOrthogonalWav, Unser:SplineWavelets}, whose scaling functions are B-splines (see \ref{sec:appendixA}).

In order to compute ${\bm{\Theta}^T \bm{x}}$, we have to determine the transposes of all the matrices in \eqref{eq:ForwardModel}. The transpose of the SRM is straightforward, since the normalized WHT matrix is unitary. The $\bm{R}$ matrix has the transpose that is again a separable two-dimensional convolution with the same sequence $r[k]$. However, in the forward model $\bm{\Theta x}$, we compute only parts of the convolution that are calculated without zero-padded edges, and in the transpose, we compute the full convolution with zero-padded edges. The transpose of $\bm{\Psi}$ is implemented by using the fast DWT algorithm with the same synthesis filters used in the IDWT. By using the proposed fast matrix-free realizations of the SRM, $\bm{R}$ and DWT, as well as of their transposes, computing a single mfIPM iteration is not exceeding $\mathcal{O}(n\log n)$. 

The mfIPM solver uses a preconditioned conjugate gradient method, in that it takes $\mathcal{O}(\log n)$ mfIPM iterations to solve the minimization problem in \eqref{eq:CSrecSIspaces}. For more information on the mfIPM solver and its computational complexity, please refer to \cite{Fountoulakis:MfIPM}.

\subsection{Measurement setups for single-pixel compressive imaging}
\label{subsec:MeasSetupCI}
We conduct the proposed CS framework, with the algorithm described above, on a set of publicly available standard test images \cite{dataBase}. The downloaded images, namely \textit{cameraman, Lena, peppers, pirate, Barbara} and \textit{boat}, are of size ${512 \times 512}$. We simulate the single-pixel measurement procedure by multiplying a test image with ${512 \times 512}$ measurement masks of $\pm 1$s, such that each image pixel has its corresponding mask pixel. A single measurement $y_m$ is obtained by summing the pixels of the image modulated with the $m$-th measurement mask.

\begin{figure}
    \centering
    \includegraphics[width=0.9\columnwidth]{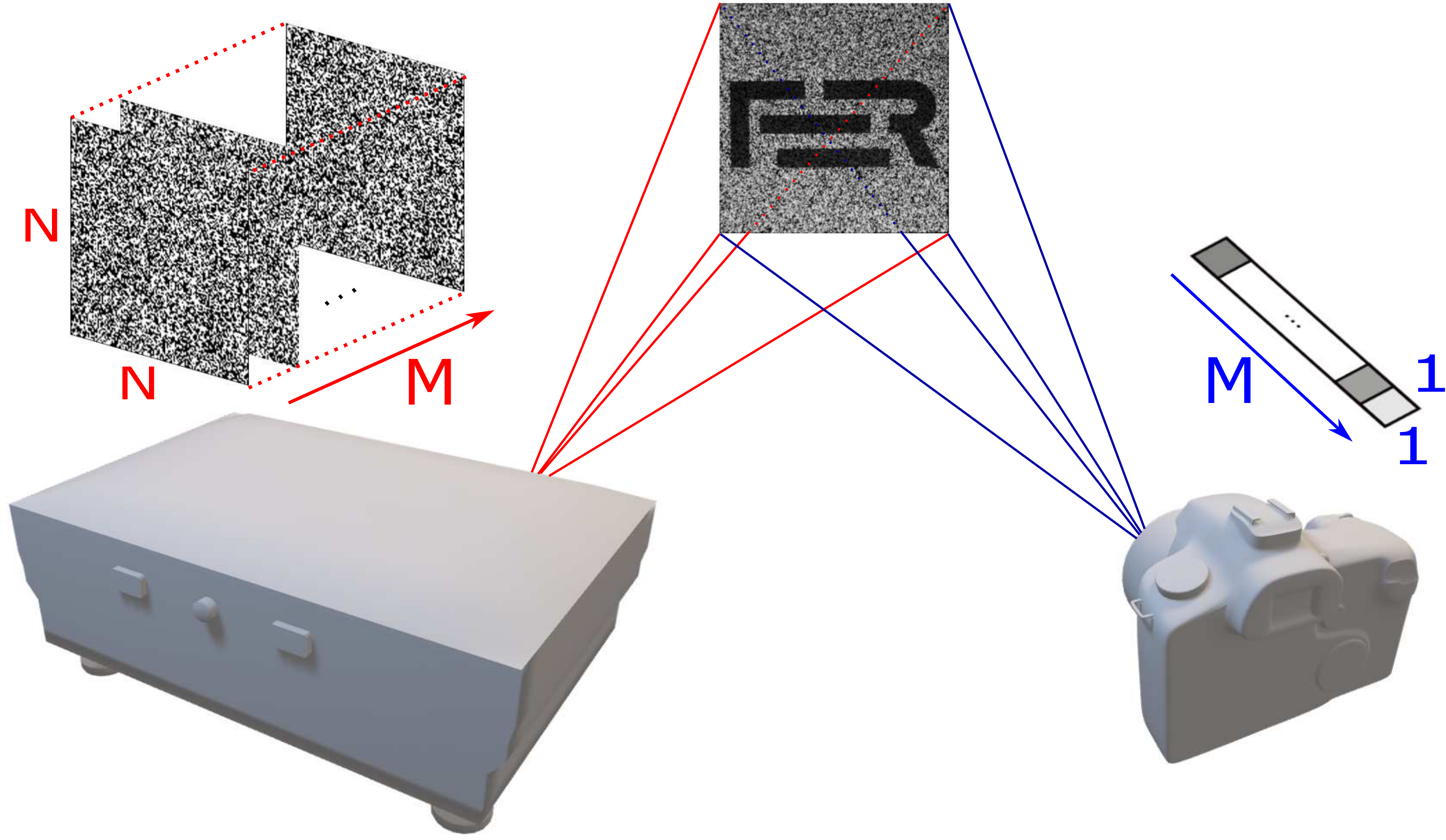}
    \caption{\textbf{Real-world imaging system scheme.} A scene is modulated by $N \times N$ patterns and $M$ measurements are obtained by the ``single-pixel" camera.}
    \label{fig:measSetup}
\end{figure}

Additionally, we test the proposed CS framework on real-world data obtained by the imaging system reported in \cite{Ralasic:OffTheShelf}. The sensing system is illustrated in Figure \ref{fig:measSetup}. It consists of a high-resolution digital projector and an off-the-shelf camera. The digital projector illuminates a scene with ${256 \times 256}$ masks and measurement data are acquired by the ``single-pixel'' camera. For low-detail scenes, good reconstruction results can be obtained from a small number of linear measurements. However, for complex scenes with a lot of details, the number of measurements is enormous if one wants to obtain high-quality reconstruction results in ${256 \times 256}$ resolution. In order to efficiently and realistically simulate measurements for a high-detail scene, we use the principle of dual photography \cite{Sen2005} and the estimated light transport matrix (LTM) between the camera and the projector by following the procedure reported in \cite{Ralasic:DualImaging}. We compute the high-resolution LTM between the projector of resolution ${256 \times 256}$ and the camera of resolution ${128 \times 128}$ with only a fraction of measurements and use it to calculate the light transport between the projector and an imaginary single-pixel detector at the camera location. Consequently, the single-pixel measurements are computed by ${\bm{y}^T = \bm{T}\bm{S}^T}$, where ${\bm{T} \in \mathbb{R}^N}$ is a light transport vector and $\bm{S}$ is an SRM of $\pm1$s. Additionally, we use high-resolution LTMs to obtain faithful ground truths of the acquired scenes.

\section{Experimental results}
\label{sec:Experiments}
We assess the image reconstruction quality of the proposed CS framework for various settings and compare it to the recovery ability of the conventionally discretized CS method. We do not compare the proposed method to the works of Eldar and Mishali \cite{Eldar:CSAnalogSI, Eldar:RobustRecovery}, and Debarre \textit{et al.}  \cite{Debarre2019, Debarre:HybridDictionaries}, which use SI subspace signal models to discretize the continuous-domain inverse problems, since the observed signals should also be sparse in the same SI subspace, and our framework is intended for reconstruction of the signals that lie in an SI subspace but, in general, are not sparse in that subspace. Additionally, we focus on compressive imaging and the way to improve its reconstruction results in continuous-domain applications by discretizing the problem exactly. We believe that the proposed discretization method can be implemented into deep reconstruction approaches for single-pixel imaging, however this is clearly beyond the scope of this paper and will be considered in the future work.

In the experiments, we use B-spline signal generators of order ${p=0,1,2,3}$ that correspond to piecewise \textit{constant}, \textit{linear}, \textit{quadratic} and \textit{cubic} splines, respectively. Note that the proposed framework matches conventional discretization of single-pixel compressive imaging for B-spline of order $0$. This setting is used mainly for the comparison of the frameworks. Note that the proposed discretization procedure is exact for all of the B-spline scaling functions, but the reconstruction depends on how well the model matches the underlying continuous-domain signal. There are various complements of the B-spline scaling functions amongst spline wavelets. Here, we employ the biorthogonal wavelets reported in \cite{Cohen:BiOrthogonalWav}, which are very popular in image coding. The biorthogonal wavelets are denoted with \textit{bior$N_r.N_d$}, where $N_r$ and $N_d$ are numbers of vanishing moments for the synthesis and analysis filters, respectively. A wavelet with $N_r$ vanishing moments is orthogonal to polynomials of degree ${N_r-1}$, and a synthesis scaling function of \textit{bior$N_r.N_d$} lies in an SI subspace spanned by the B-spline of order ${p=N_r-1}$.

\subsection{Numerical experiments}
\label{subsec:NumExperiments}
First, we conduct the proposed framework on a set of standard test images. The reconstruction quality is assessed for various numbers of measurements that are computed by the simulation of single-pixel imaging described in Section \ref{subsec:MeasSetupCI}. We define a measurement ratio ${m_r=M/N}$ as the ratio between the number of measurement masks $M$ and the number of pixels ${N=KL}$ in the masks, and select ten numbers $M$ that correspond to the ratios ${m_r = 5\%, 10\%, \dots, 50\%}$. Same measurement masks are used for different orders $p$ of the B-spline basis functions in order to reject the impact of a random matrix realization on the reconstruction ability. Image reconstruction is obtained by solving the proposed $\ell_1$-minimization problem in \eqref{eq:CSrecSIspaces}. The regularization parameter $\lambda$ is determined experimentally for every setting, such that it leads to the best reconstruction results in terms of the peak signal-to-noise ratio (PSNR). Reconstruction quality is assessed in terms of the PSNR in decibels and structural similarity index measure (SSIM).

\begin{figure*}[t!]
	\centering
		\subfloat{%
        	\includegraphics[width=0.32\textwidth]{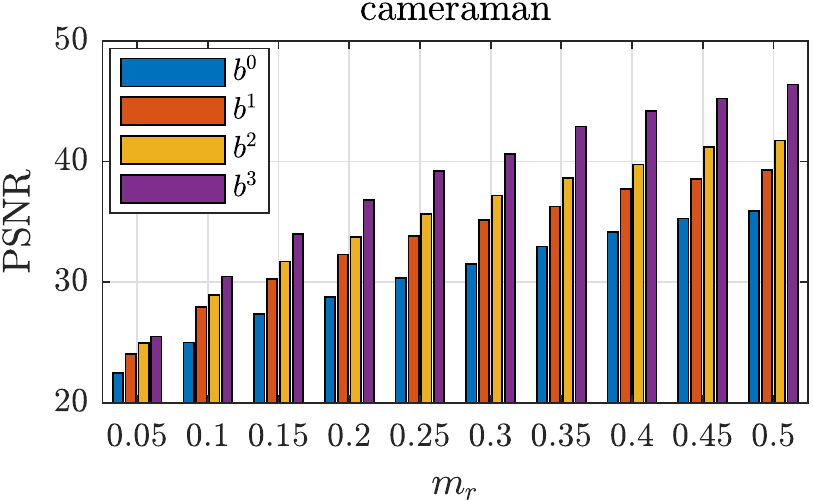}%
        }
        \hspace{1ex}
		\subfloat{%
			\includegraphics[width=0.32\textwidth]{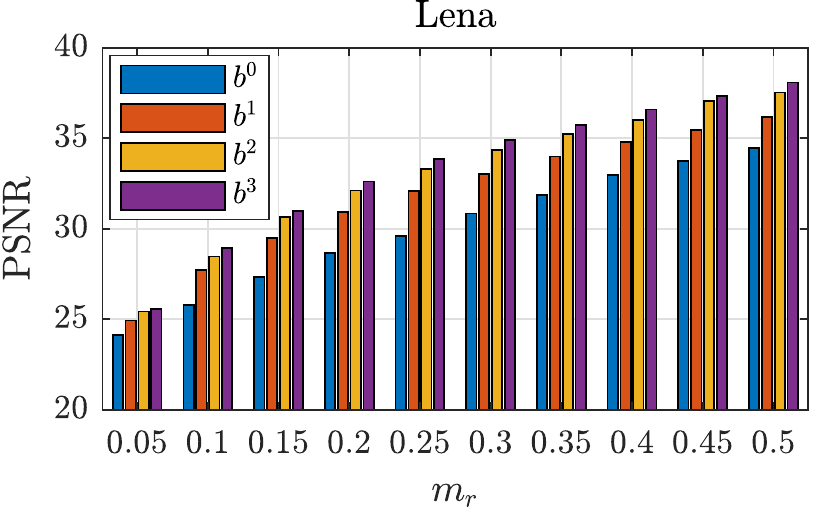}%
        }
        \hspace{1ex}
		\subfloat{%
			\includegraphics[width=0.32\textwidth]{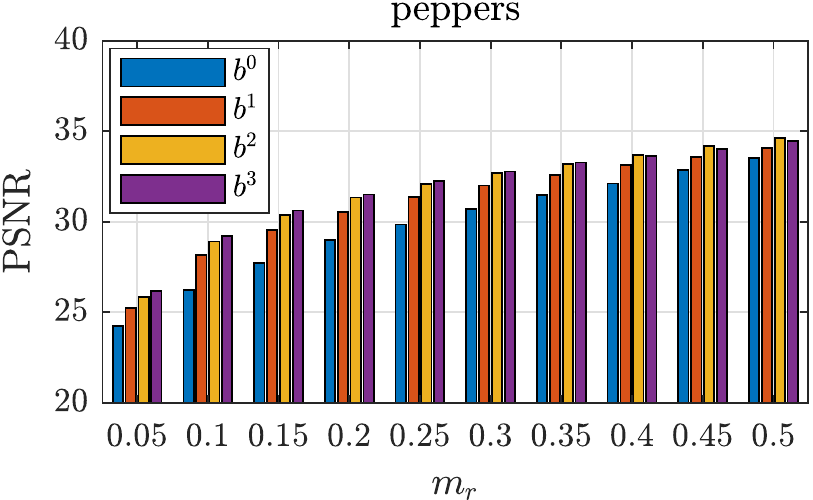}%
        }
        \vspace{-1.2ex}
        
		\subfloat{%
        	\includegraphics[width=0.32\textwidth]{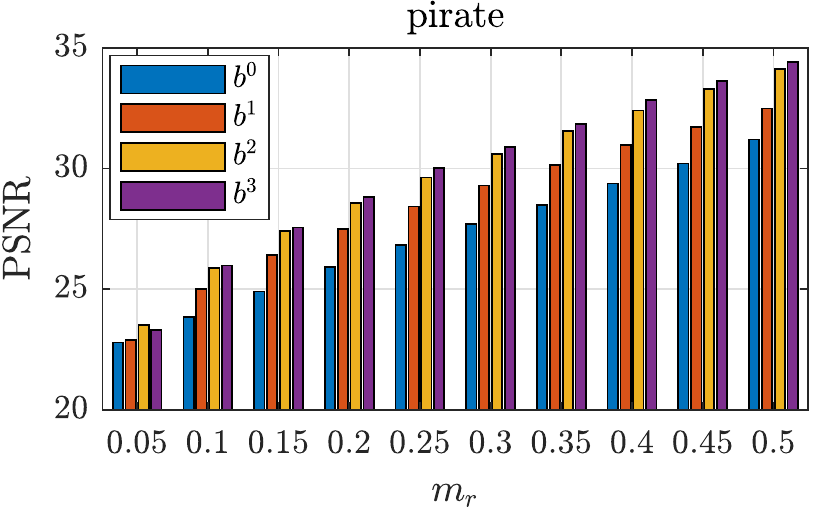}%
        }
        \hspace{1ex}
		\subfloat{%
			\includegraphics[width=0.32\textwidth]{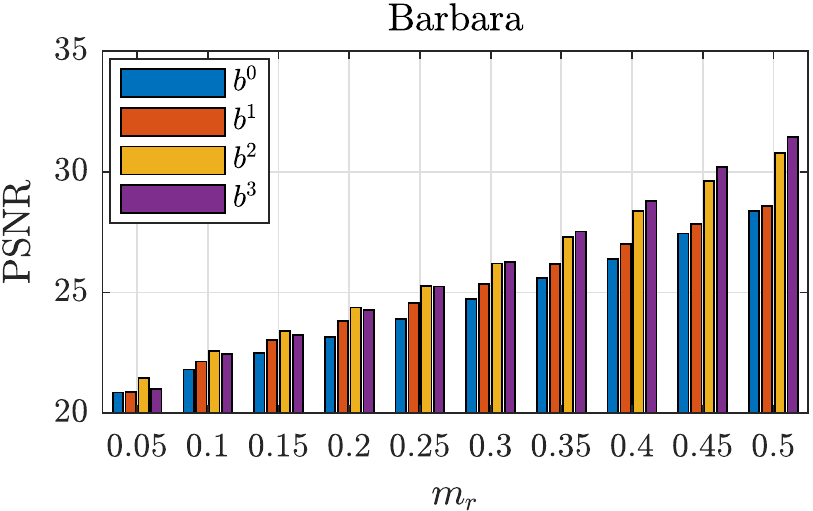}%
        }
        \hspace{1ex}
		\subfloat{%
			\includegraphics[width=0.32\textwidth]{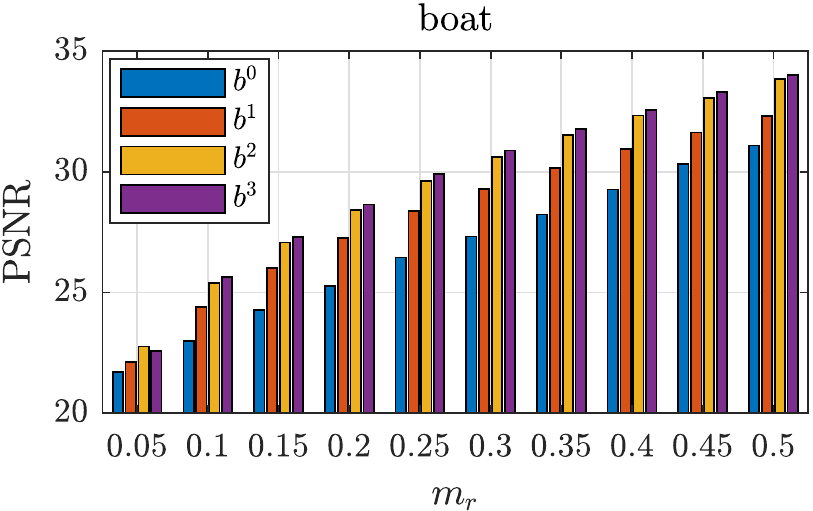}%
        }
    \vspace{-1.55ex}
	\caption{\textbf{Reconstruction results for the \textit{bior2.2} wavelets in terms of the PSNR.} For a single measurement ratio $m_r$, there are four bars representing reconstruction results consecutively for the B-spline approximation functions of order $0$ to $3$. Results of the conventional discretization method correspond to the first (blue) bar in each group.}
	\vspace{-1.25ex}
	\label{fig:chartBior2.2}
\end{figure*}

\begin{figure*}[t!]
	\centering
		\subfloat{%
        	\includegraphics[width=0.3225\textwidth]{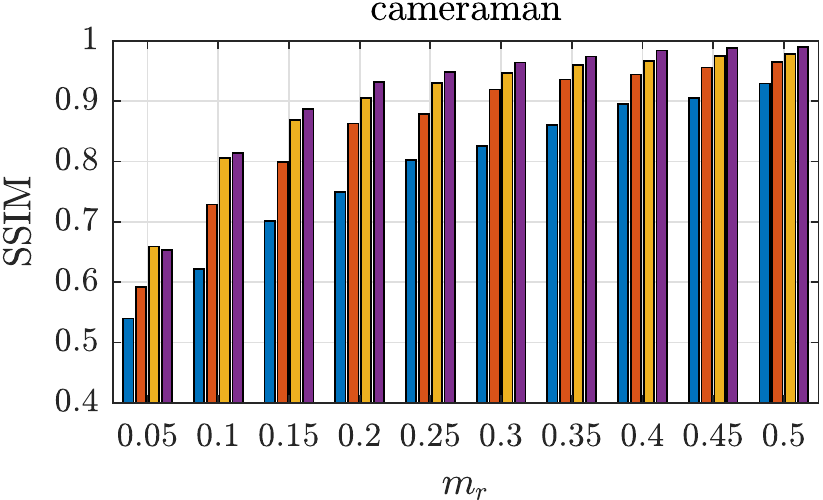}%
        }
        \hspace{1ex}
		\subfloat{%
			\includegraphics[width=0.3225\textwidth]{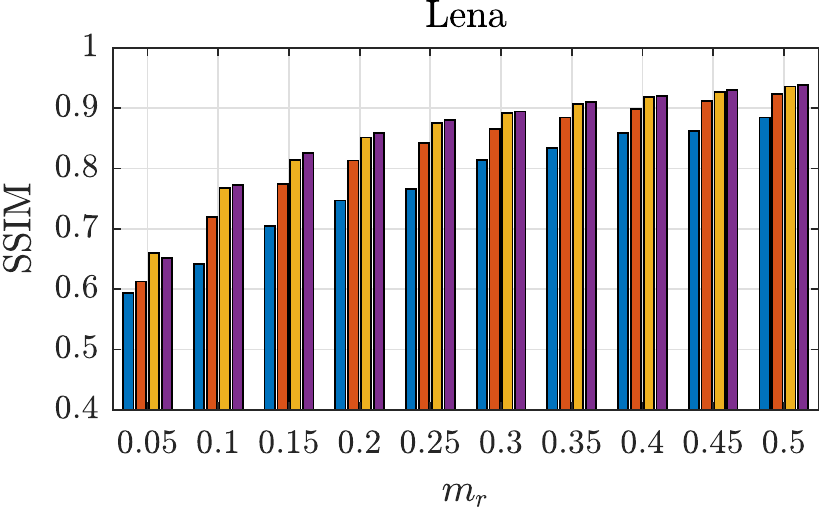}%
        }
        \hspace{1ex}
		\subfloat{%
			\includegraphics[width=0.3225\textwidth]{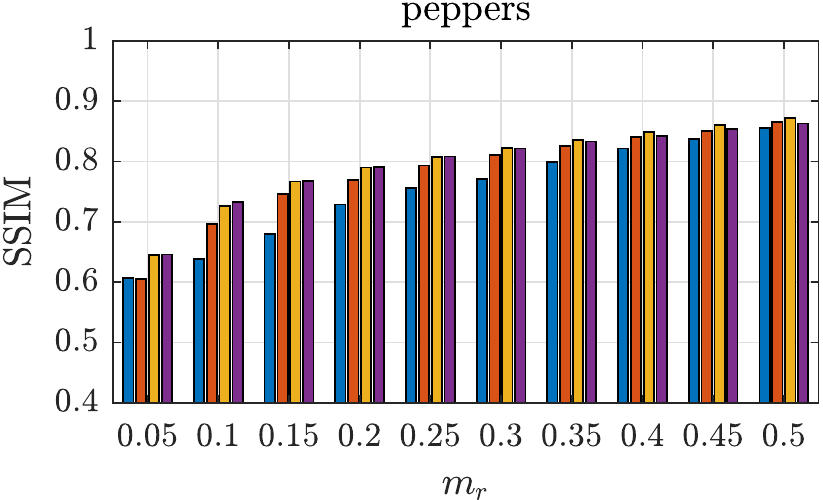}%
        }
        \vspace{-1.2ex}
        
		\subfloat{%
        	\includegraphics[width=0.3225\textwidth]{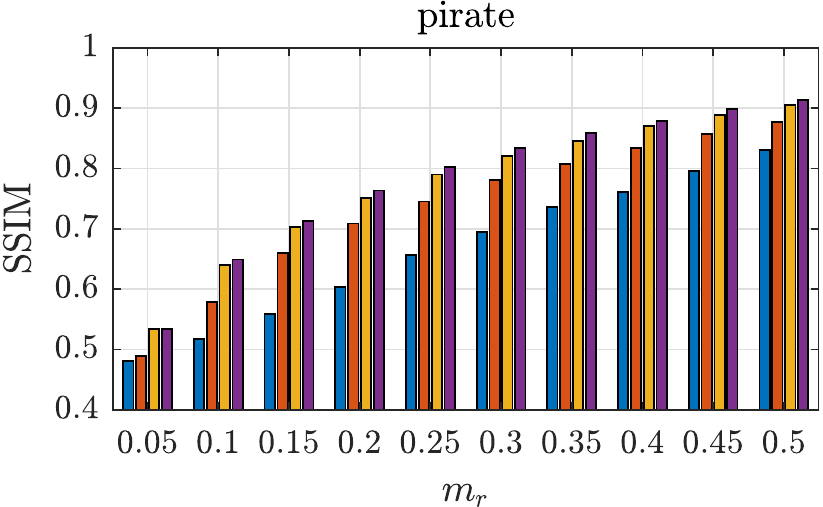}%
        }
        \hspace{1ex}
		\subfloat{%
			\includegraphics[width=0.3225\textwidth]{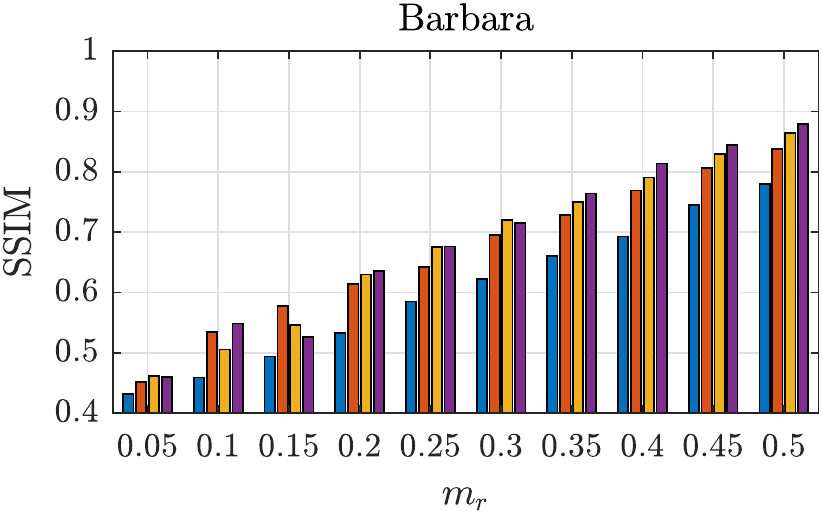}%
        }
        \hspace{1ex}
		\subfloat{%
			\includegraphics[width=0.3225\textwidth]{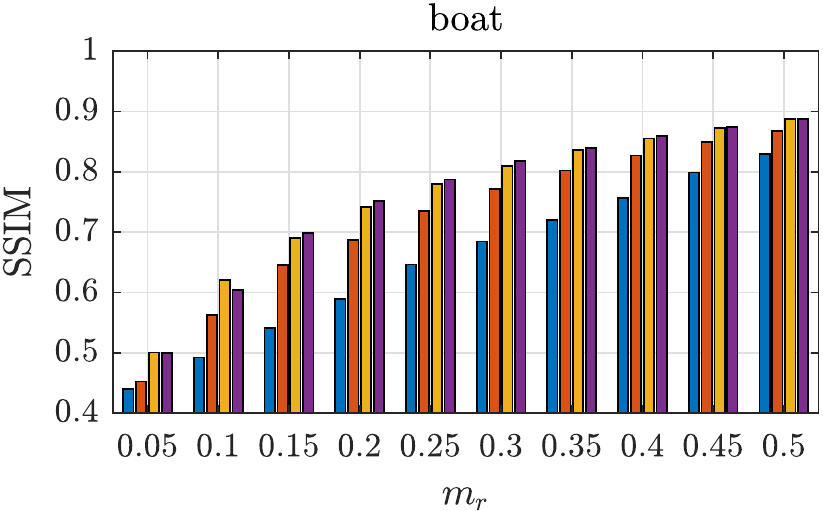}%
        }
    \vspace{-1.55ex}
	\caption{\textbf{Reconstruction results for the \textit{bior2.2} wavelets in terms of the SSIM.} For a single measurement ratio $m_r$, there are four bars representing reconstruction results consecutively for the B-spline approximation functions of order $0$ to $3$. Results of the conventional discretization method correspond to the first (blue) bar in each group. The color legend is the same as is in Figure \ref{fig:chartBior2.2}.}
	\vspace{-1.25ex}
	\label{fig:SSIMchartBior2.2}
\end{figure*}

\begin{figure*}[t]
	\centering
		\subfloat{%
        	\includegraphics[width=0.32\textwidth]{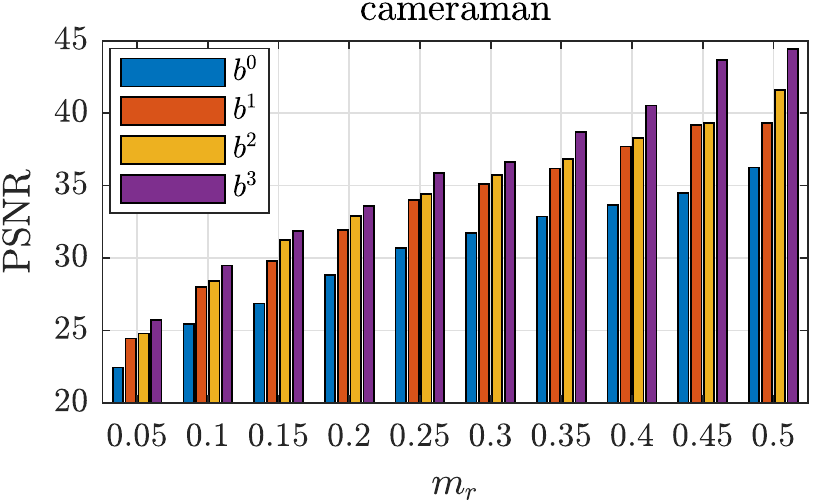}%
        }
        \hspace{1ex}
		\subfloat{%
			\includegraphics[width=0.32\textwidth]{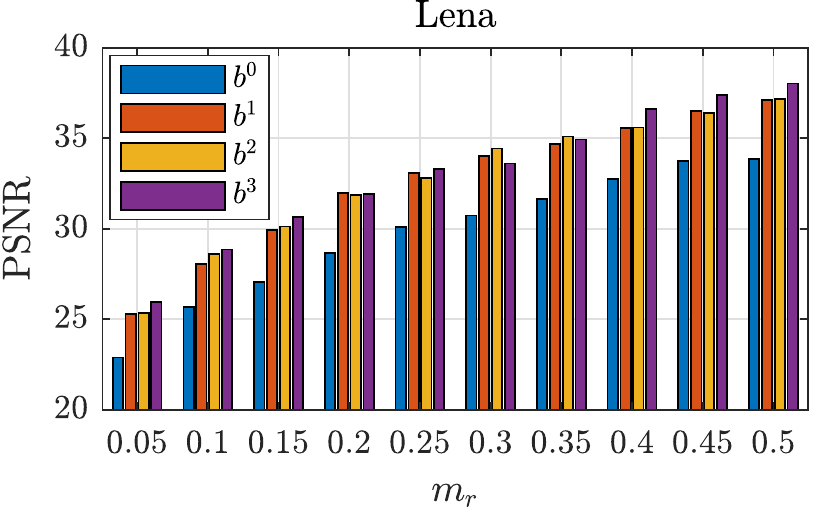}%
        }
        \hspace{1ex}
		\subfloat{%
			\includegraphics[width=0.32\textwidth]{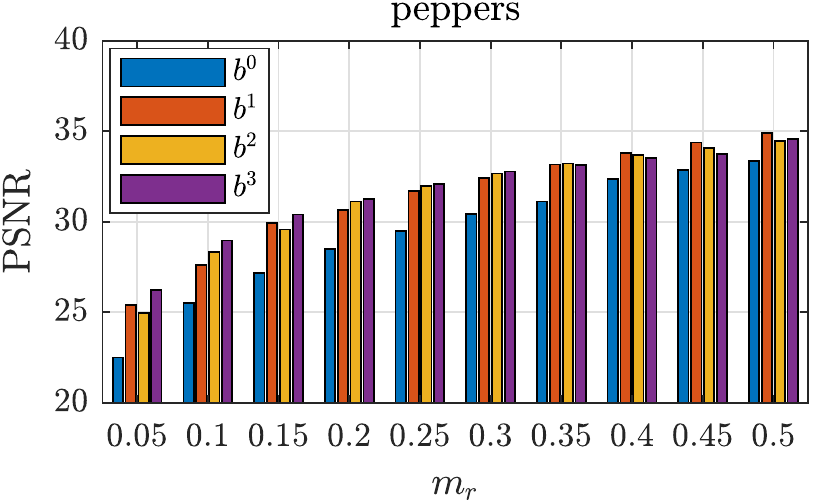}%
        }
        \vspace{-1.2ex}
        
		\subfloat{%
        	\includegraphics[width=0.32\textwidth]{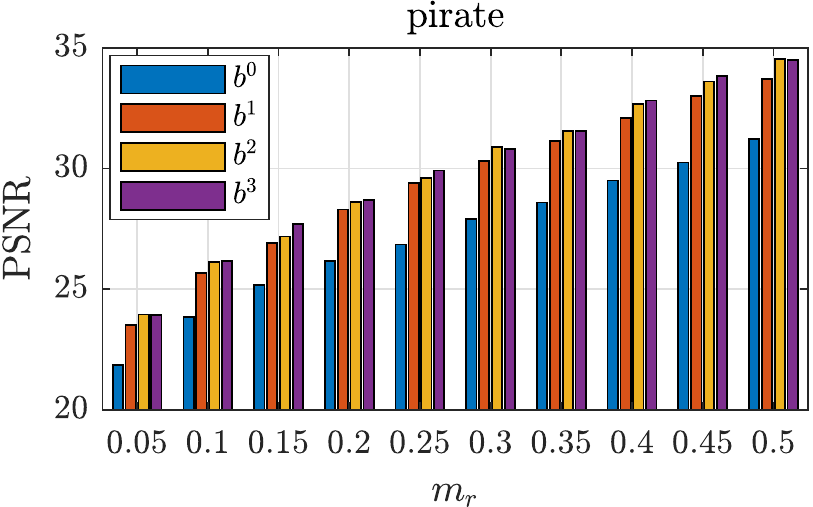}%
        }
        \hspace{1ex}
		\subfloat{%
			\includegraphics[width=0.32\textwidth]{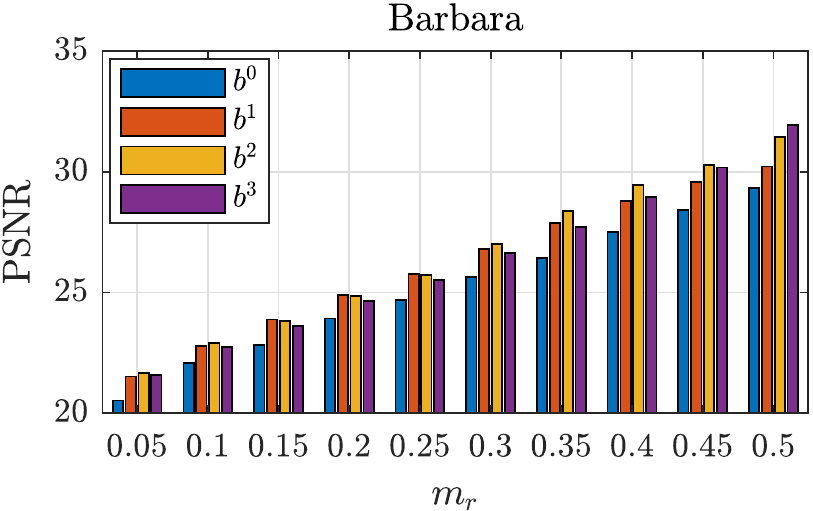}%
        }
        \hspace{1ex}
		\subfloat{%
			\includegraphics[width=0.32\textwidth]{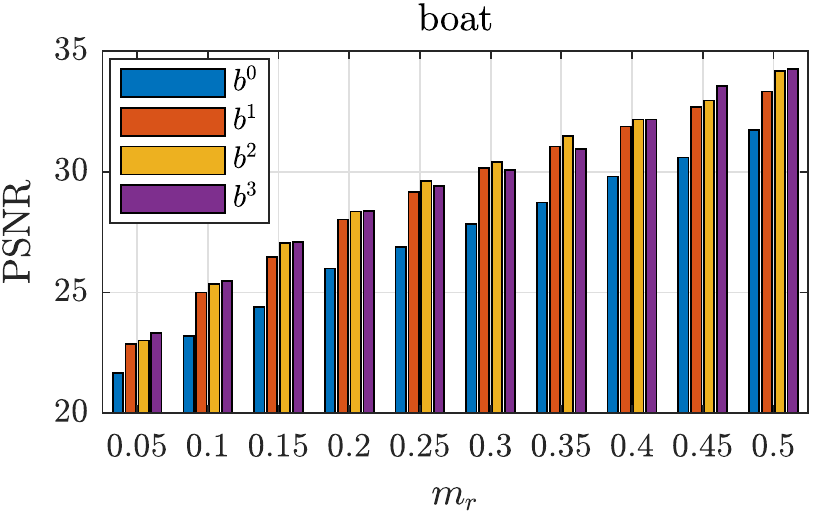}%
        }
    \vspace{-1.55ex}
	\caption{\textbf{Reconstruction results for the \textit{bior4.4} wavelets in terms of the PSNR.} For a single measurement ratio $m_r$, there are four bars representing reconstruction results consecutively for the B-spline approximation functions of order $0$ to $3$. Results of the conventional discretization method correspond to the first (blue) bar in each group.}
	\vspace{-1.65ex}
	\label{fig:chartBior4.4}
\end{figure*}

\begin{figure*}[t!]
	\centering
		\subfloat{%
        	\includegraphics[width=0.3225\textwidth]{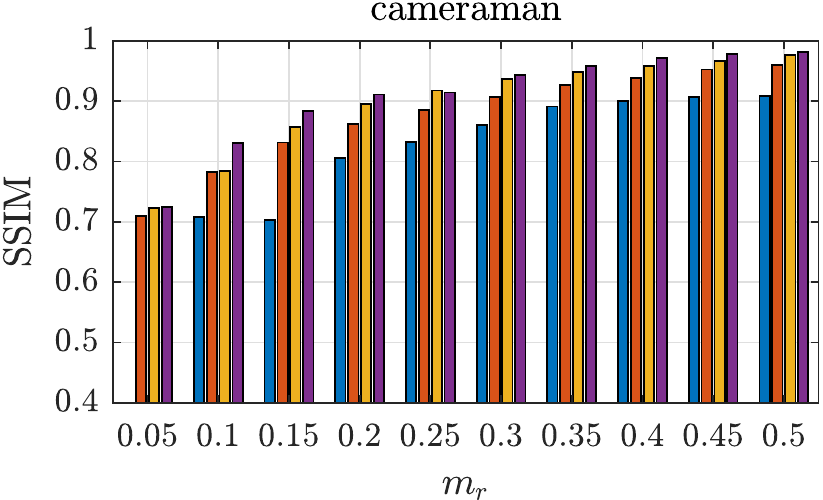}%
        }
        \hspace{1ex}
		\subfloat{%
			\includegraphics[width=0.3225\textwidth]{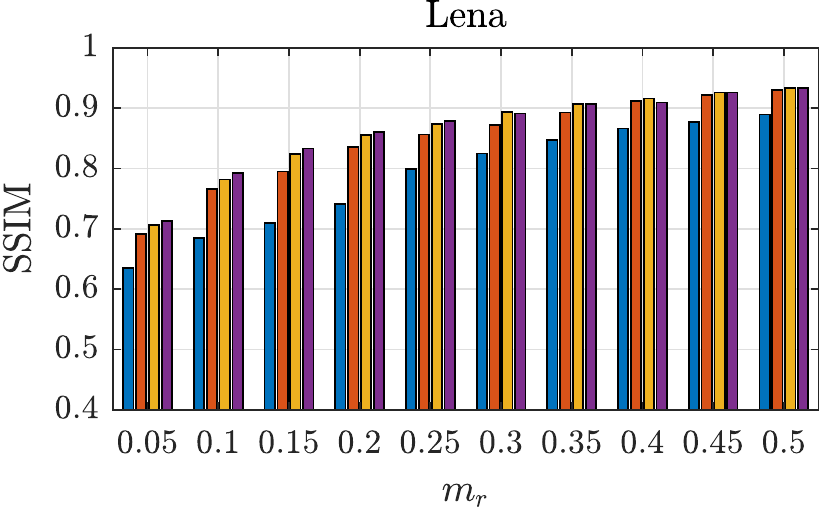}%
        }
        \hspace{1ex}
		\subfloat{%
			\includegraphics[width=0.3225\textwidth]{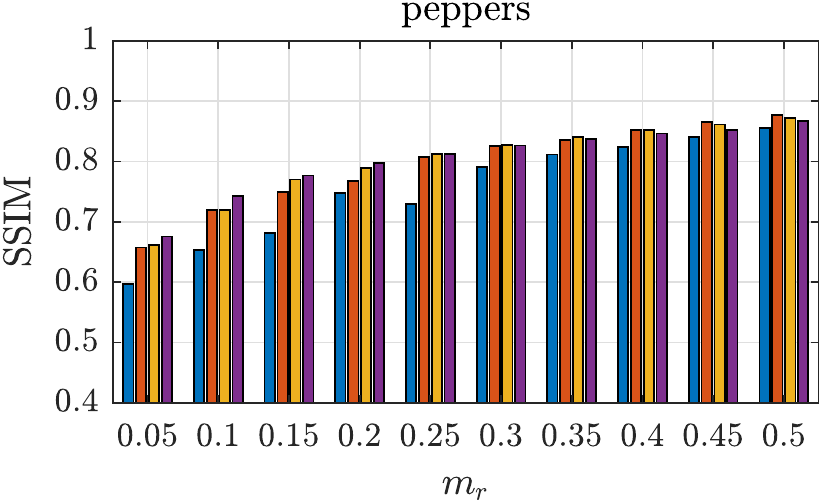}%
        }
        \vspace{-1.2ex}
        
		\subfloat{%
        	\includegraphics[width=0.3225\textwidth]{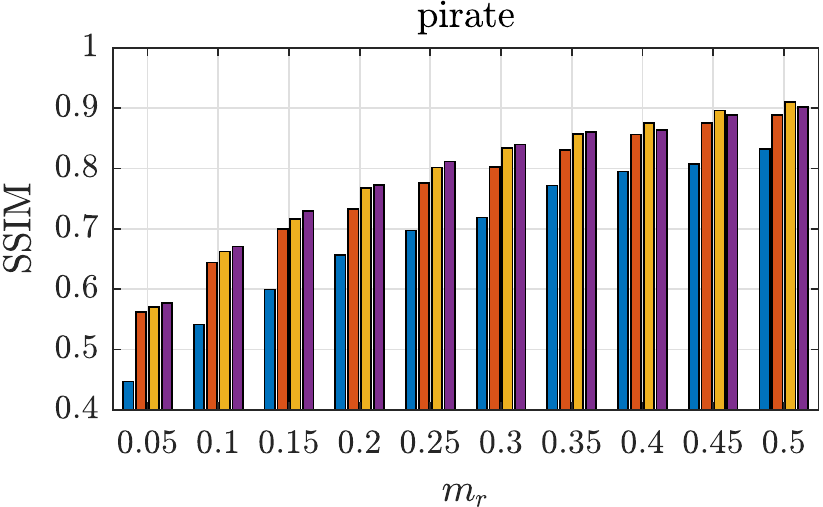}%
        }
        \hspace{1ex}
		\subfloat{%
			\includegraphics[width=0.3225\textwidth]{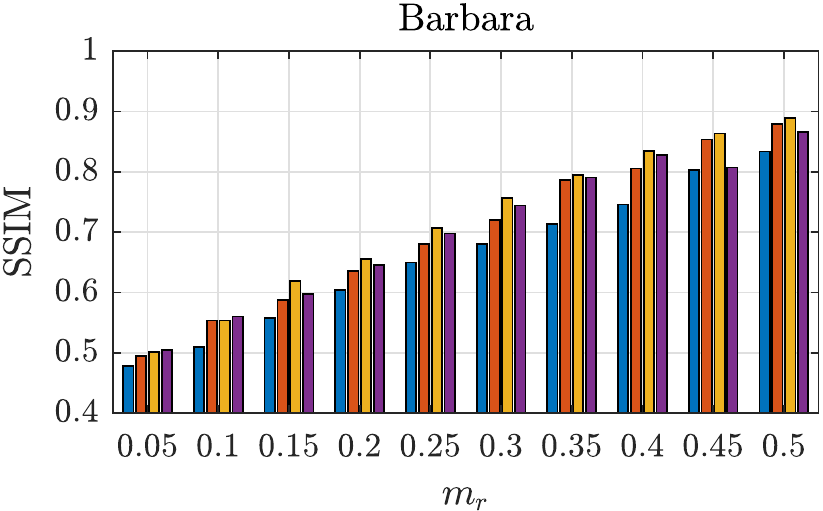}%
        }
        \hspace{1ex}
		\subfloat{%
			\includegraphics[width=0.3225\textwidth]{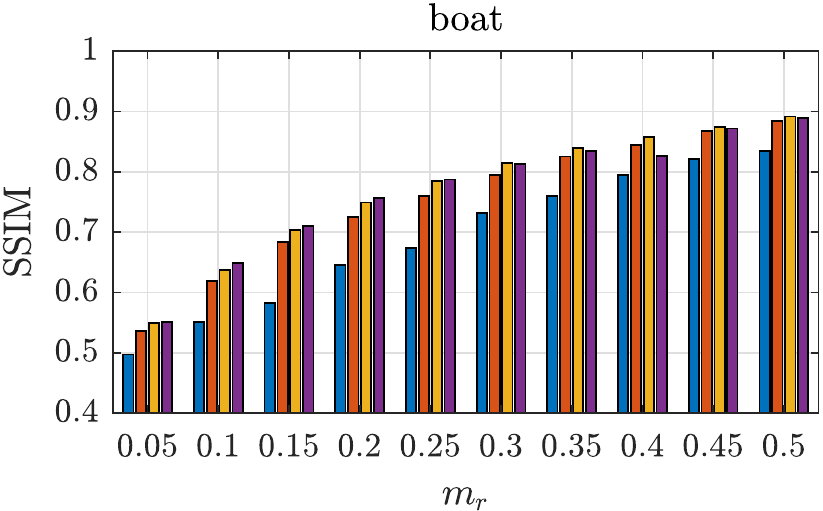}%
        }
    \vspace{-1.55ex}
	\caption{\textbf{Reconstruction results for the \textit{bior4.4} wavelets in terms of the SSIM.} For a single measurement ratio $m_r$, there are four bars representing reconstruction results consecutively for the B-spline approximation functions of order $0$ to $3$. Results of the conventional discretization method correspond to the first (blue) bar in each group. The color legend is the same as is in Figure \ref{fig:chartBior4.4}.}
	\vspace{-1.65ex}
	\label{fig:SSIMchartBior4.4}
\end{figure*}

Reconstruction results of the test images for various settings and the \textit{bior2.2} wavelets are given by the bar charts in Figure \ref{fig:chartBior2.2} and Figure \ref{fig:SSIMchartBior2.2}.
%
The scaling function for \textit{bior2.2} is the B-spline of order $1$. When the approximation function ${\varphi(t)=b^1(t)}$, the inverse problem is exactly discretized in a sense that an input signal is decomposed by the continuous-domain wavelet functions and the recovered vector $\bm{x}$ contains the coefficients given in the two-dimensional version of \eqref{eq:WaveletTransform}. However, we are not solely limited to use the approximation function that corresponds to the scaling function of a chosen wavelet. The approximation coefficients $\bm{a}_{(0)}$ are a sequence of numbers that can be considered sparse in various transform domains. The proposed discretization method is still exact, which is secured by the $\bm{R}$ matrix, and \eqref{eq:ApproxCoeffsAndSparsityMtx} holds, where $\bm{x}$ are coefficients of $\bm{a}_{(0)}$ in a certain discrete transform basis $\bm{\Psi}$ (\textit{e.g.}, discrete cosine transform). Accordingly, for other B-spline basis functions, the approximation coefficients $\bm{a}_{(0)}$ can also be assumed as sparse in the same wavelet domain. However, $\bm{x}$ obtained in this way does not represent the coefficients that are obtained by the two-dimensional wavelet decomposition \eqref{eq:WaveletTransform}. The same can be applied to the \textit{bior4.4} wavelets whose reconstruction results are given in Figure \ref{fig:chartBior4.4} and Figure \ref{fig:SSIMchartBior4.4}.
%

In both Figure \ref{fig:chartBior2.2} and Figure \ref{fig:chartBior4.4}, from the charts, it can be seen that the B-spline approximation functions $b^1(t)$, $b^2(t)$ and $b^3(t)$ achieve better results in terms of the PSNR than $b^0(t)$, which is associated with the conventional discretization of the CS problem. The results in Figure \ref{fig:SSIMchartBior2.2} and Figure \ref{fig:SSIMchartBior4.4} show that the proposed discretization framework enhances the reconstruction quality in terms of the SSIM in a similar manner in comparison with the conventional discretization in the CS setting. Even though $b^0(t)$ is orthogonal to the sampling kernel ${\zeta(t)=b^0(t)}$ and leads to the simplified CS problem where ${\bm{R}:=\bm{I}}$, it has proven to be too coarse an approximation function in the majority single-pixel compressive imaging setups. That is, we can not expect that the observed continuous-domain signal is adequately modeled as a piecewise constant signal, but rather it lies in some space that is better approximated by B-splines of higher orders. In most settings, the $b^3(t)$ approximation function yields the best reconstruction results, achieving the same reconstruction qualities as $b^0(t)$ for approximately two times lower measurement ratios. The proposed framework achieves outstanding results for the \textit{cameraman} when $b^3(t)$ is used, which makes us suspect that its downloaded version was previously interpolated possibly with the cubic spline.

\begin{figure}[t]
	\centering
		\subfloat[Ground truth]{%
        	\includegraphics[width=0.23\textwidth]{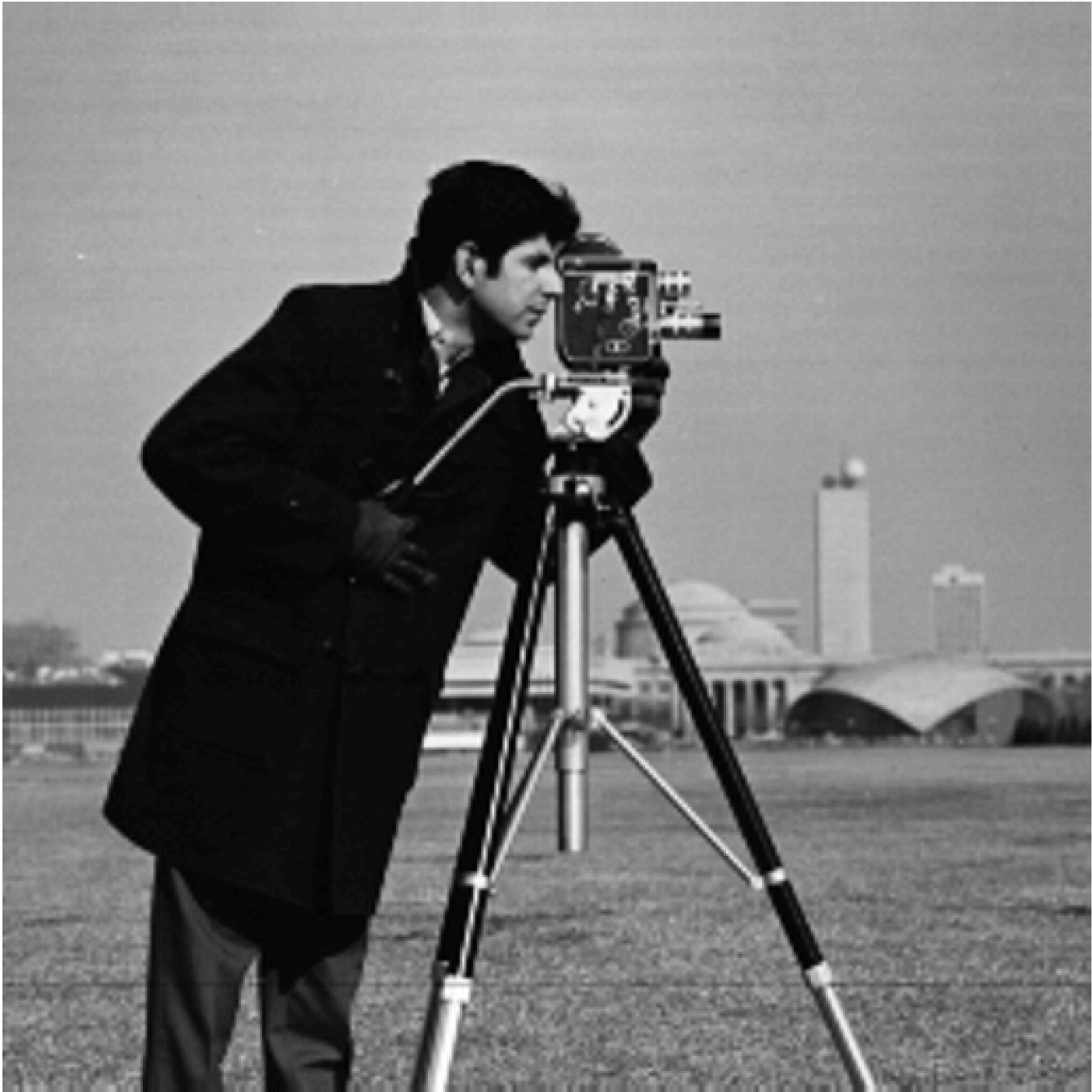}%
        }
        \hspace{1ex}
		\subfloat[$b^0(t)$, PSNR: ${30.03 \, \mathrm{dB}}$]{%
			\includegraphics[width=0.23\textwidth]{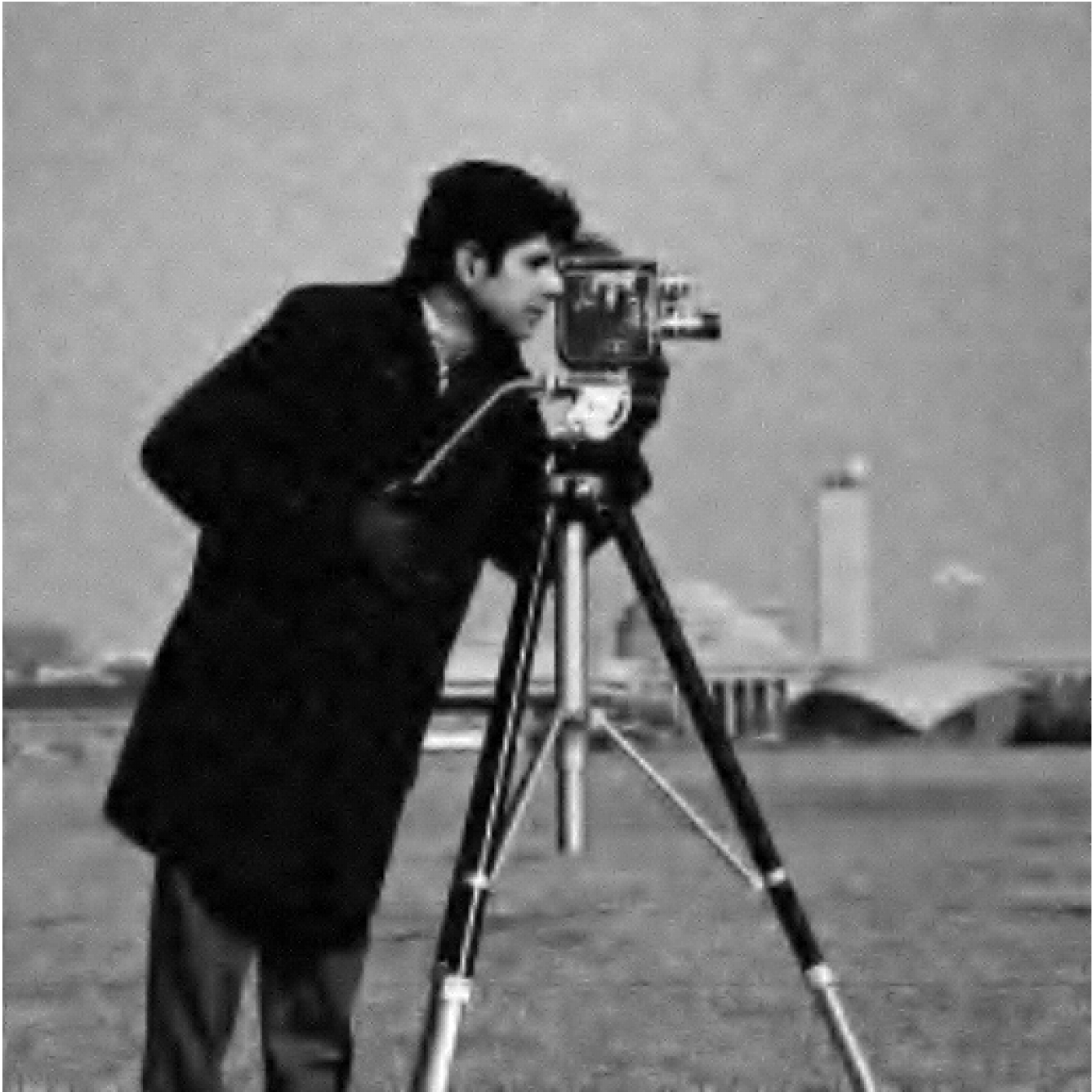}%
        }
        \vspace{-1ex}
        
		\subfloat[$b^1(t)$, PSNR: ${33.82 \, \mathrm{dB}}$]{%
			\includegraphics[width=0.23\textwidth]{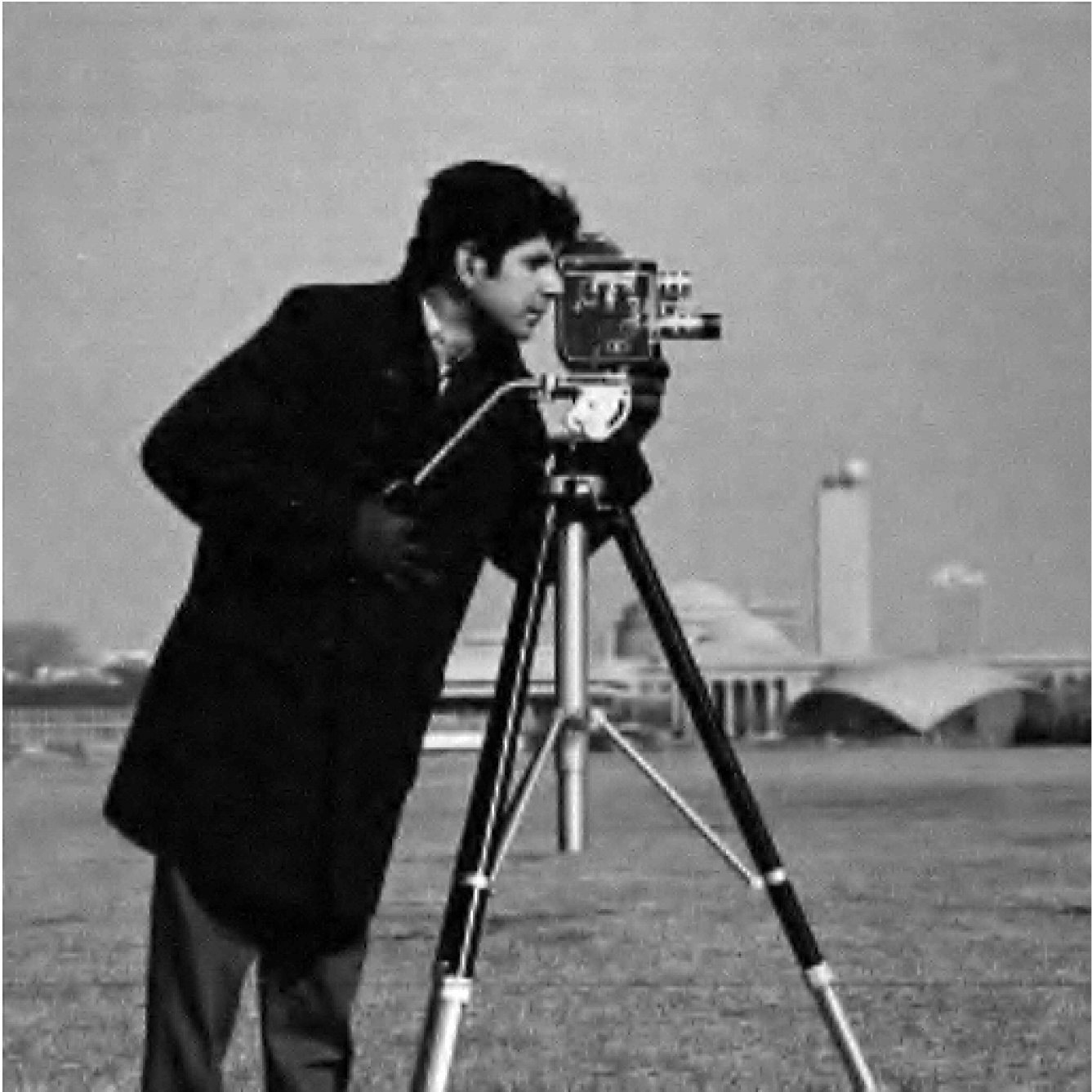}%
        }
        \hspace{1ex}
		\subfloat[$b^3(t)$, PSNR: ${38.17 \, \mathrm{dB}}$]{%
			\includegraphics[width=0.23\textwidth]{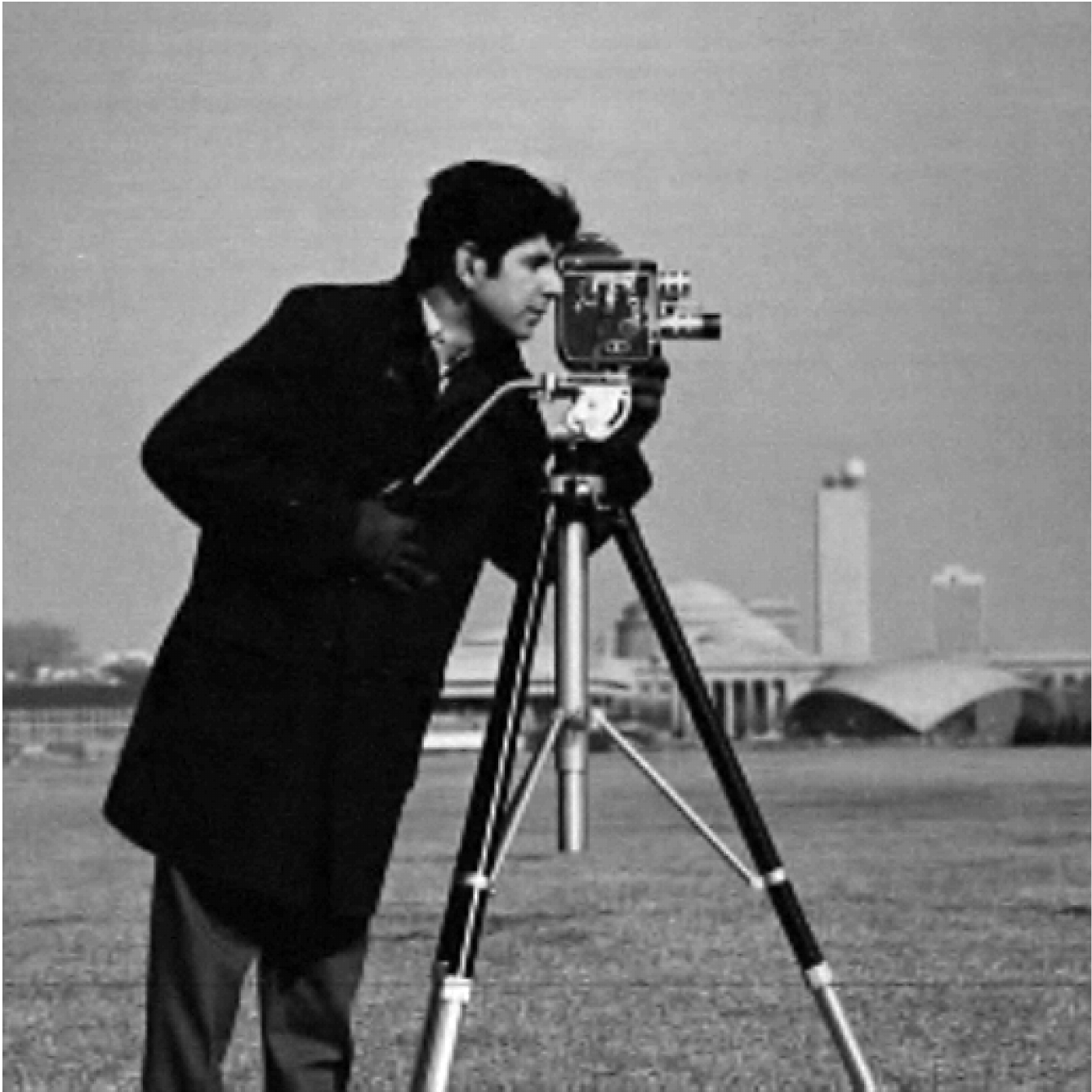}%
        }
	\caption{\textbf{Reconstructions of the \textit{cameraman} image for the B-spline approximation functions of order $\mathbf{0, 1}$ and $\mathbf{3}$.} The ground truth and reconstructions are of ${512 \times 512}$ size. Measurement ratio is set to ${m_r = 25\%}$ and sparsity basis is spanned by the \textit{bior2.2} wavelets. Note that the reconstruction in (b) corresponds to the conventional discretization method.}
	\vspace{-1.5ex}
	\label{fig:cameraman}
\end{figure}

\begin{figure}[t]
	\centering
		\subfloat[Ground truth]{%
        	\includegraphics[width=0.23\textwidth]{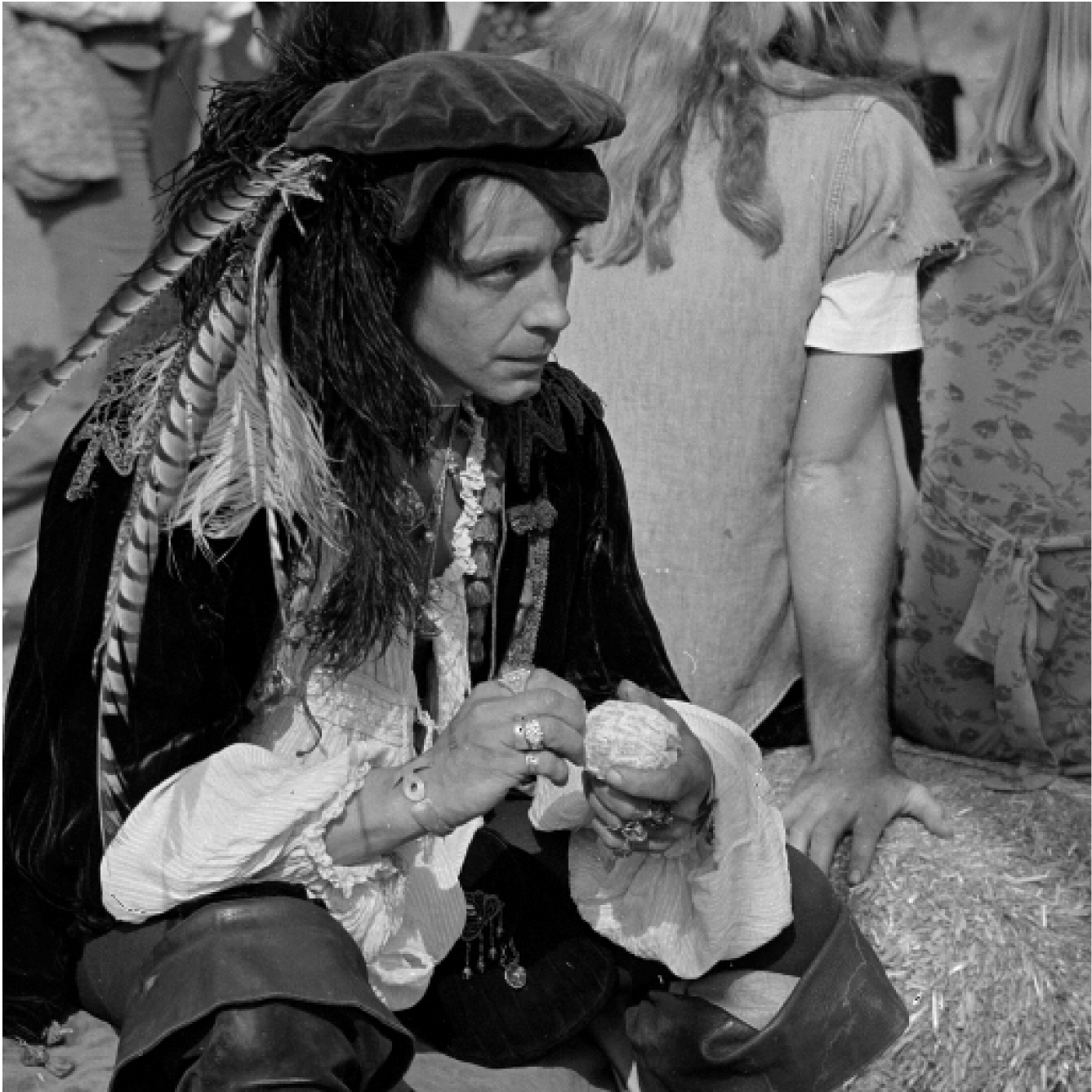}%
        }
        \hspace{1ex}
		\subfloat[$b^0(t)$, PSNR: ${26.85 \, \mathrm{dB}}$]{%
			\includegraphics[width=0.23\textwidth]{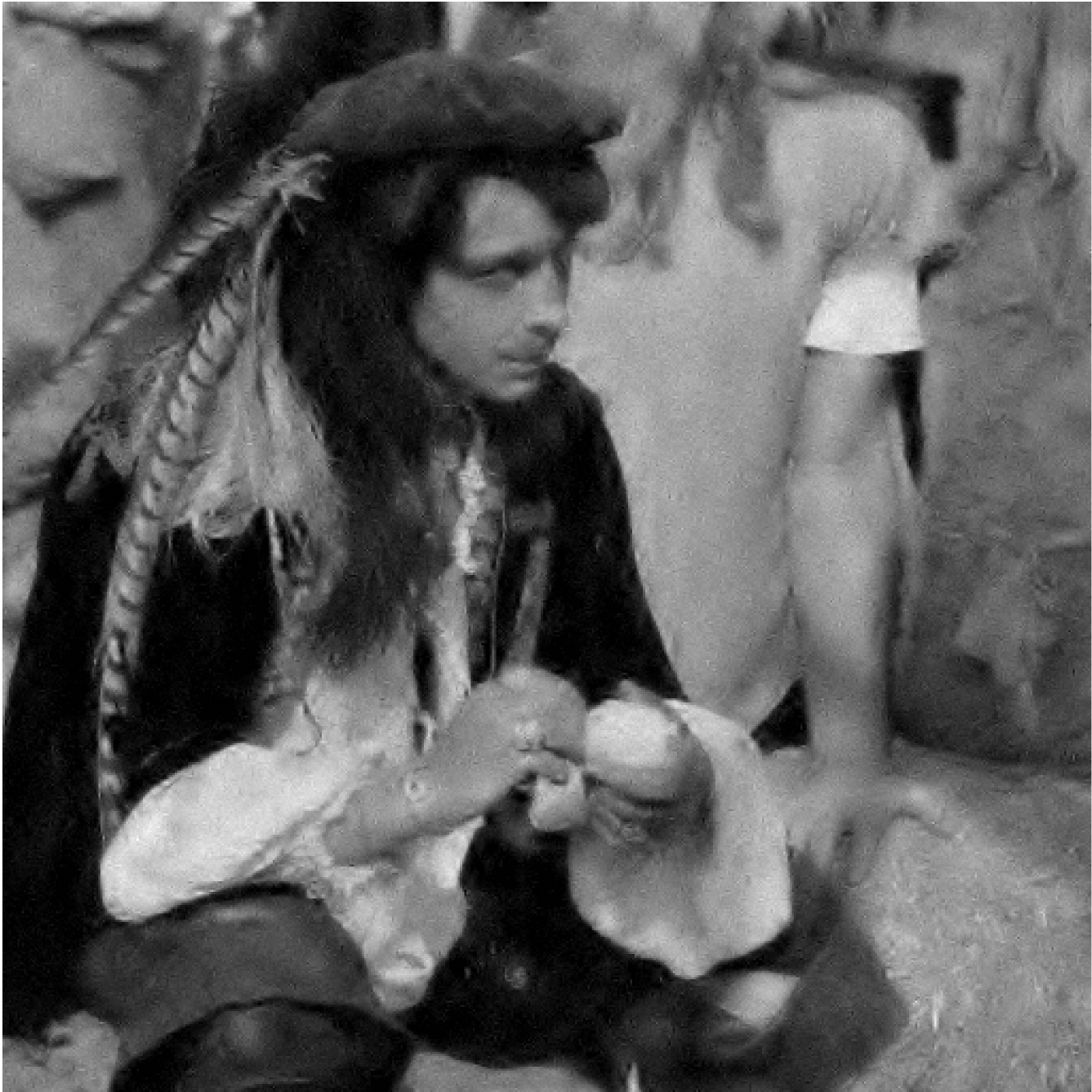}%
        }
        \vspace{-1ex}
        
		\subfloat[$b^1(t)$, PSNR: ${29.38 \, \mathrm{dB}}$]{%
			\includegraphics[width=0.23\textwidth]{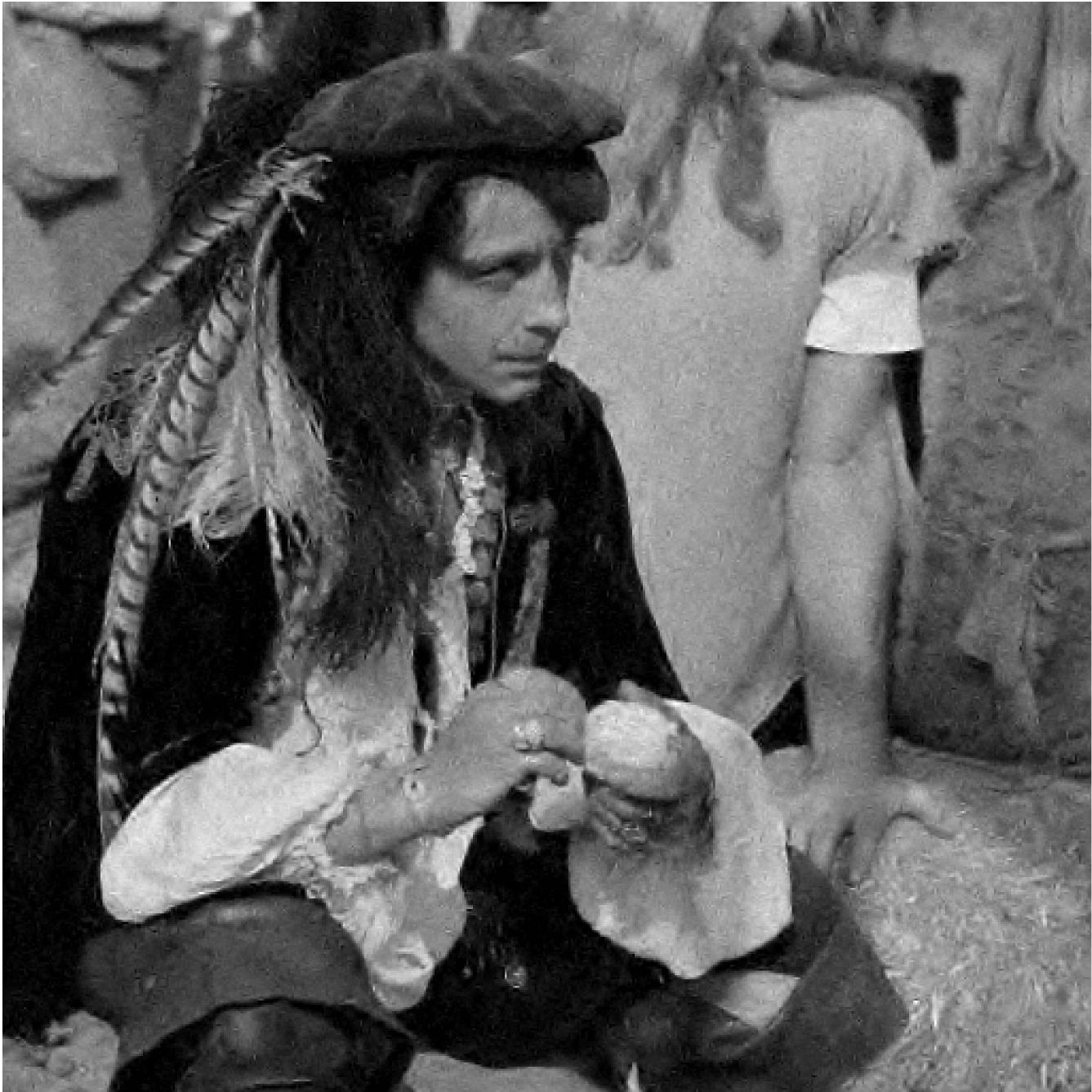}%
        }
        \hspace{1ex}
		\subfloat[$b^3(t)$, PSNR: ${29.91 \, \mathrm{dB}}$]{%
			\includegraphics[width=0.23\textwidth]{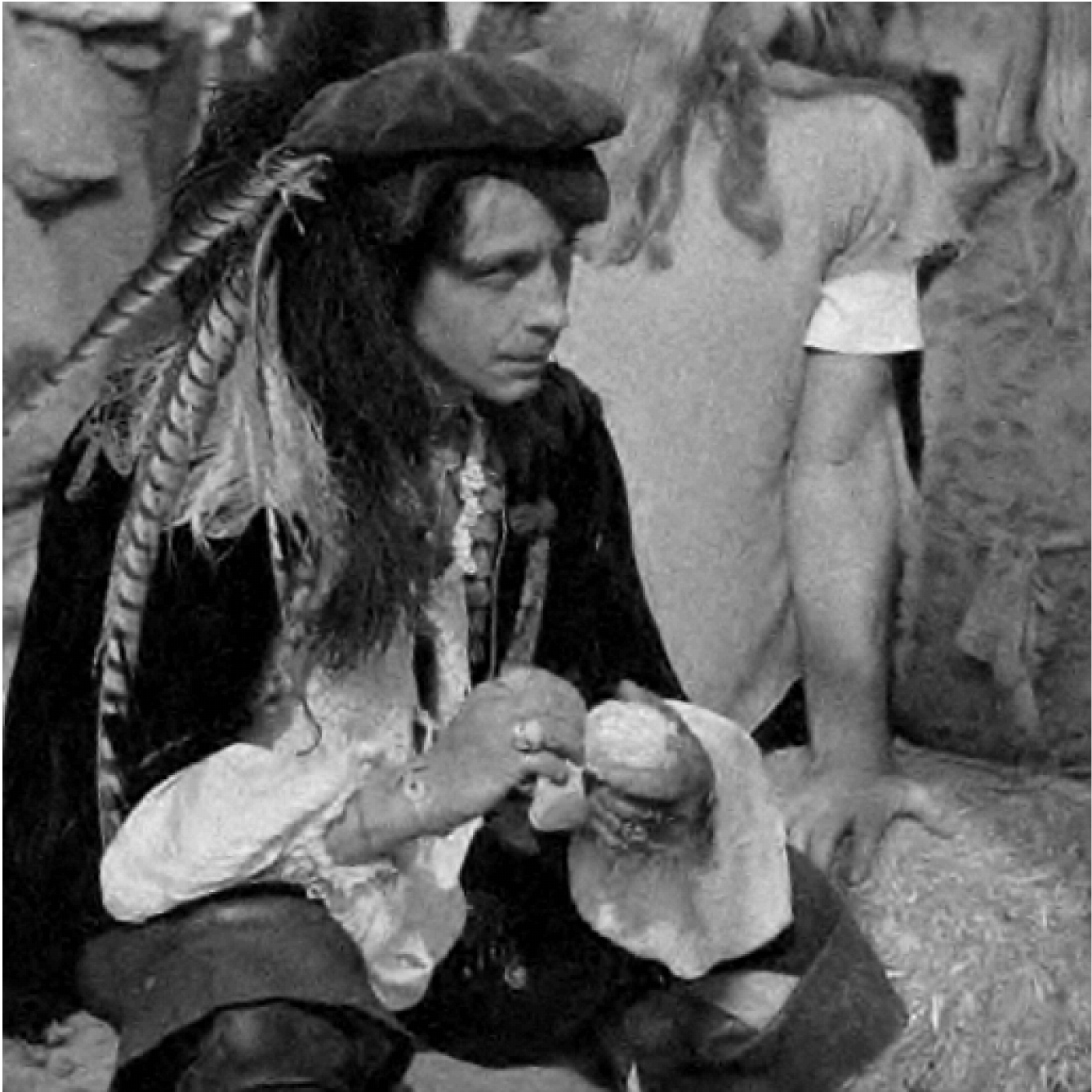}%
        }
	\caption{\textbf{Reconstructions of the \textit{pirate} image for the B-spline approximation functions of order $\mathbf{0, 1}$ and $\mathbf{3}$.} The ground truth and reconstructions are of ${512 \times 512}$ size. Measurement ratio is set to ${m_r = 25\%}$ and sparsity basis is spanned by the \textit{bior4.4} wavelets. Note that the reconstruction in (b) corresponds to the conventional discretization method.}
	\vspace{-1ex}
	\label{fig:pirate}
\end{figure}

Figure \ref{fig:cameraman} and Figure \ref{fig:pirate}
%
show single image reconstructions of the \textit{cameraman} and \textit{pirate} for two different settings. The enhancement of the reconstruction quality for the proposed discretization method of single-pixel compressive imaging relative to the conventional one is clearly visible.

\subsection{Real-world data experiments}
\label{subsec:RWDataExp}
\begin{figure}[t]
	\centering
		\subfloat[Ground truth]{%
        	\includegraphics[width=0.23\textwidth]{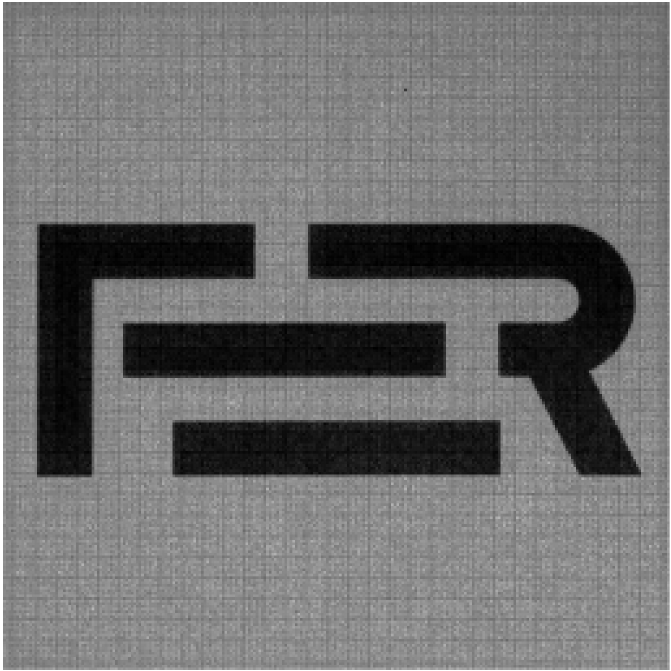}%
        }
        \hspace{1ex}
		\subfloat[$b^0(t)$, SSIM: $0.0840$]{%
			\includegraphics[width=0.23\textwidth]{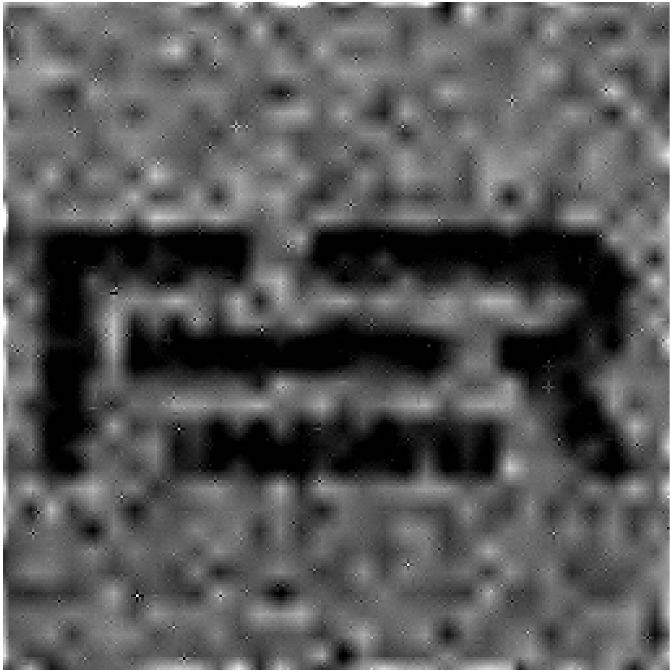}%
        }
        \vspace{-1.5ex}
        
		\subfloat[$b^3(t)$, SSIM: $0.0995$]{%
			\includegraphics[width=0.23\textwidth]{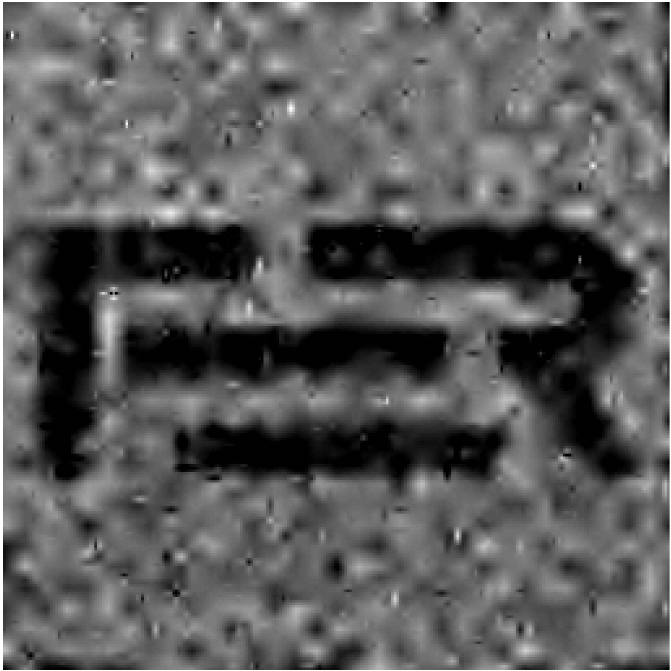}%
        }
        \hspace{1ex}
		\subfloat[$b^5(t)$, SSIM: $0.1103$]{%
			\includegraphics[width=0.23\textwidth]{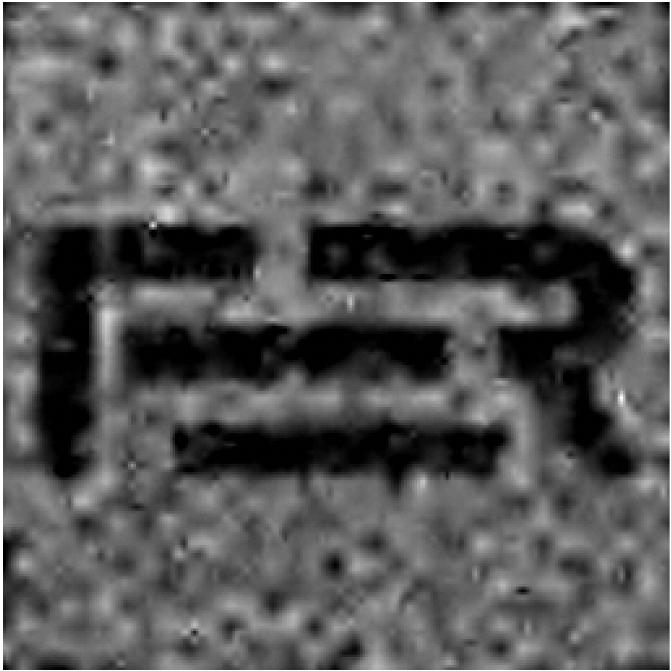}%
        }        
	\caption{\textbf{Reconstructions of a scene acquired by a real compressive imaging system.} B-spline approximation functions are of order $0$, $3$ and $5$. Measurement ratio is set to ${m_r=12.5\%}$. The ground truth is a \textit{dual image} from the point of view of the projector which is obtained by dual photography. Note that the reconstruction in (b) corresponds to the conventional discretization method.}
	\vspace{-1.5ex}
	\label{fig:RWMD_ferLogo}
\end{figure}

Figure \ref{fig:RWMD_ferLogo} shows reconstruction results of a low-detail scene by applying the proposed CS framework on real-world data obtained by the compressive imaging system described in Section \ref{subsec:MeasSetupCI} and Figure \ref{fig:measSetup}. Real-world imaging applications are highly corrupted by additive and multiplicative noise, which both affect reconstruction results in Figure \ref{fig:RWMD_ferLogo}. The projector illuminated $8,192$ patterns of $256 \times 256$ resolution and the modulated scene was captured by the ``single-pixel" camera. The regularization parameter $\lambda$ was tuned specifically for every setting to yield the highest SSIM. The proposed CS discretization method outperforms the conventional one in terms of the reconstruction quality. This demonstrates the robustness of the proposed method in real-world applications which are expected to be noise corrupted.

\begin{figure}[t]
	\centering
		\subfloat[Ground truth]{%
        	\includegraphics[width=0.23\textwidth]{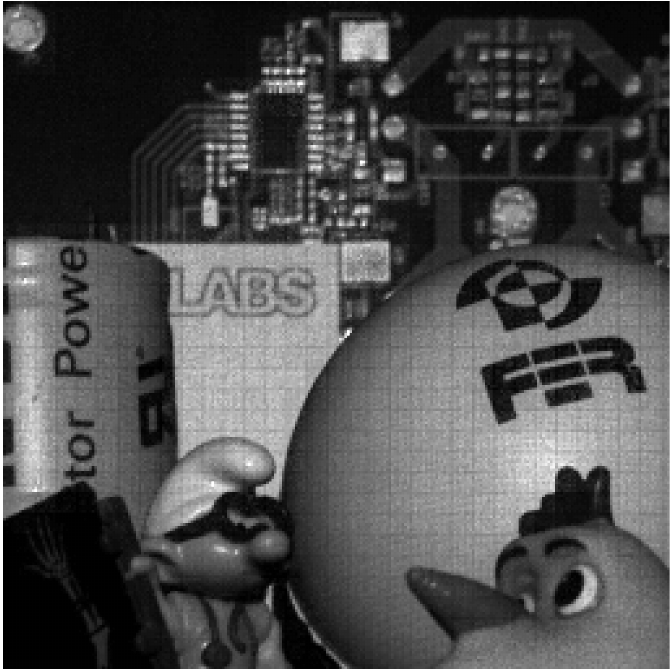}%
        }
        \hspace{1ex}
		\subfloat[$b^0(t)$, SSIM: $0.4375$]{%
			\includegraphics[width=0.23\textwidth]{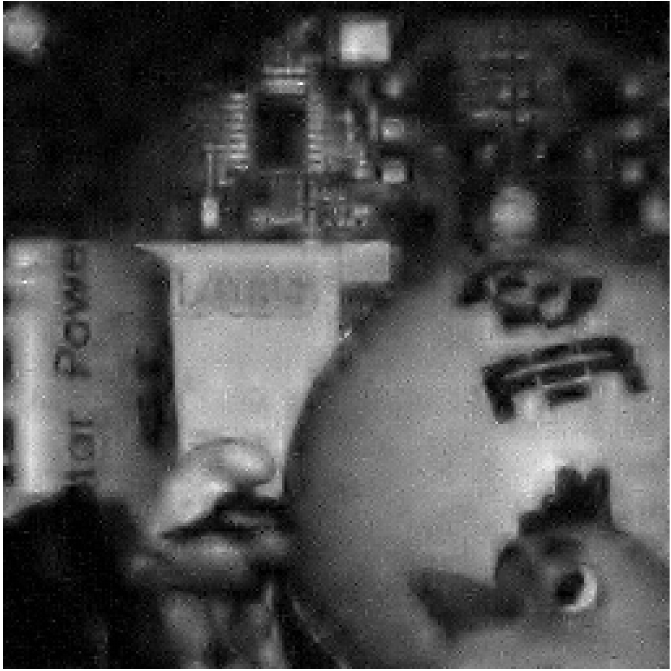}%
        }
        \vspace{-1.5ex}
        
		\subfloat[$b^1(t)$, SSIM: $0.5628$]{%
			\includegraphics[width=0.23\textwidth]{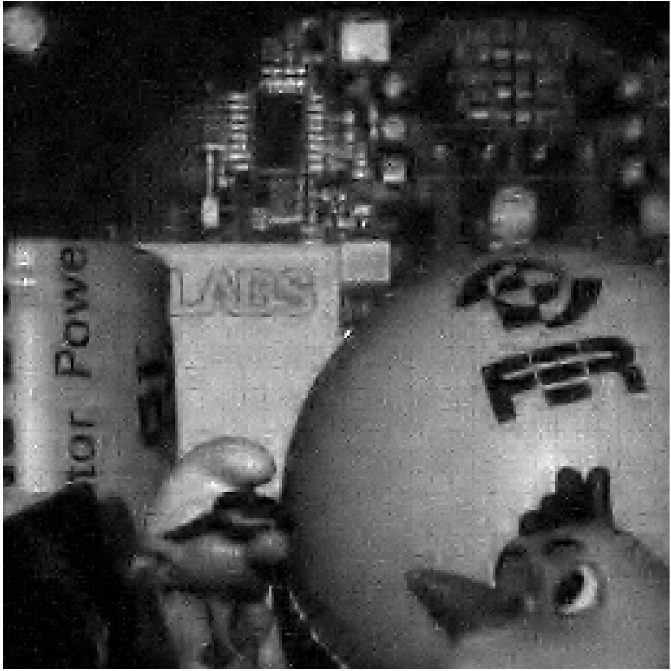}%
        }
        \hspace{1ex}
		\subfloat[$b^3(t)$, SSIM: $0.6137$]{%
			\includegraphics[width=0.23\textwidth]{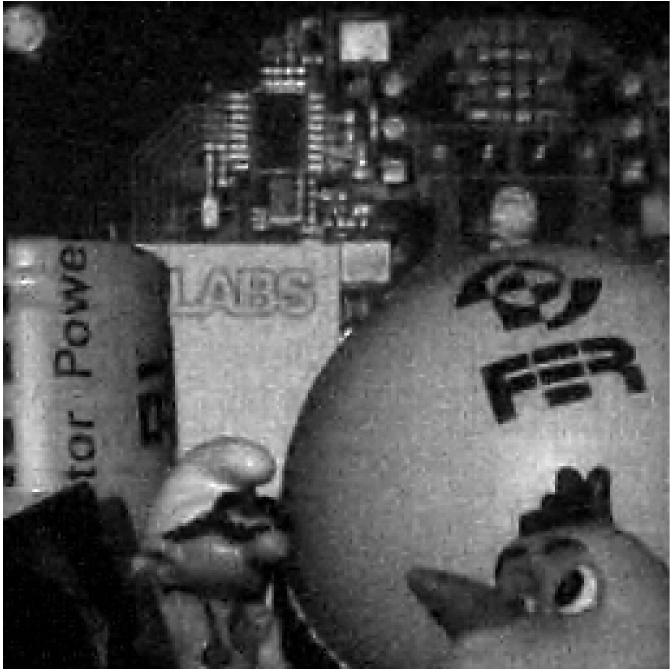}%
        }        
	\caption{\textbf{Reconstructions of a scene acquired by a dual imaging system.} B-spline approximation functions are of order $0$, $1$ and $3$. Measurement ratio is set to ${m_r=30\%}$. The ground truth is a \textit{dual image} from the point of view of the projector which is obtained by dual photography. Note that the reconstruction in (b) corresponds to the conventional discretization method.}
	\vspace{-1.5ex}
	\label{fig:DI_smurf}
\end{figure}

For reconstruction of a high-detail scene, we used the LTM to compute the measurements of an imaginary single-pixel camera (see Section \ref{subsec:MeasSetupCI}). The LTM was estimated from $1,024$ CS measurements. For more details on estimating the LTM by using a reduced set of measurements, please refer to \cite{Ralasic:DualImaging}. We simulated $19,660$ single-pixel measurements without multiplicative noise and reconstructed images are shown in Figure \ref{fig:DI_smurf}. Again, $\lambda$ was tuned such that it leads to the best reconstruction results for every setting in terms of the SSIM. Both the B-spline representations of order $1$ and $3$ yield high-quality reconstruction results and overachieve the conventional discretization characterized by the B-spline of order $0$. In all settings in both Figure \ref{fig:RWMD_ferLogo} and Figure \ref{fig:DI_smurf}, we used the \textit{bior2.2} wavelets.

\section{Related work}
\label{sec:RelatedWork}
In \cite{Eldar:CSAnalogSI} and \cite{Eldar:RobustRecovery}, analog signals are assumed to lie in unions of SI subspaces. An SI signal model is used to exactly discretize a continuous-domain inverse problem and to induce sparsity which is modeled by assuming that only a few out of all generators are active. Compressive sensing is employed in order to reduce the number of analog filters preceding sampling at the rate of innovation. The framework can be extended to a special case of sampling of analog signals that lie in an SI space spanned by a single generator. However, this leads to a signal model whose expansion coefficients $\bm{a}_0$ are assumed sparse in a sense that only a few of them are nonzero. The authors propose analog filters with sampling kernels that are linear combinations of the shifts of a function biorthogonal to the generator. Such filters may be quite difficult to implement in hardware, especially in imaging modalities, and are often approximated which introduces errors in the system.

Unser \textit{et al.} \cite{Unser:SplinesUnSolution} introduce a method for solving continuous-domain inverse problems with sparse spline solutions by using the total-variation regularization. As demonstrated in \cite{Debarre2019, Debarre:HybridDictionaries, Bohra2020}, which are papers that rely on the main results of \cite{Unser:SplinesUnSolution}, the grid-based discretization strategies lead to convex optimization problems that can be efficiently solved. The framework exploits an assumption that the innovation occurs only at a few locations out of all on some predefined grid. The signal model can be interpreted as a spline that has a small number of nonzero coefficients $\bm{a}_0$.

In \cite{Vlasic:CSinSIspaces}, the authors propose an SI signal model for an exact discretization of one-dimensional continuous-time inverse problems. The underlying signal is assumed to lie in a single SI subspace, but contrarily to the works mentioned above, the SI model does not induce sparsity. Instead, it introduces an alternative approach to the Nyquist-Shannon signal model, which is conventionally assumed in CS. In order to employ CS, the authors assume that the expansion coefficients, which are not necessarily pointwise values of the signal, are sparse in a certain discrete transform domain. The paper reinterprets the random demodulator \cite{Tropp:BeyondNyquist} as a system for CS of signals in more general class of SI subspaces. The emphasis is put on the generalization of the system to arbitrary sampling kernels and conditions on the boundaries of integration intervals.

In this paper, we use an SI signal model to recast the two-dimensional continuous-domain inverse problem of single-pixel imaging as a finite dimensional CS problem in an exact way. The emphasis is put on the multiresolution analysis that yields complement sparsity bases to the B-spline representation functions. When a representation function corresponds to a certain wavelet scaling function, the problem of signal decomposition is exact. Additionally, we offer an efficient implementation that can be used for reconstruction of large-scale images. Finally, extensive experiments were conducted on the standard test images and data obtained by a real acquisition system.

\section{Discussion and conclusion}
\label{sec:Conclusion}
We proposed a novel framework for single-pixel imaging via CS. The paper successfully links the theories of CS and generalized sampling in SI spaces. The SI property of the underlying continuous-domain signal model leads to an exact discretization of the ill-posed inverse problem. Such a model perfectly fits the CS paradigm, where the $\ell_1$-norm regularization promotes sparse solutions. In order to induce sparsity, we exploit the assumption that the expansion coefficients of the underlying signal are sparse in a certain sparsity domain.

Traditionally, the signal is assumed to be bandlimited, \textit{i.e.}, the underlying signal model rests on the Nyquist-Shannon principles. However, it is just an approximation in the case of single-pixel compressive imaging. We showed that the conventional problem is an exact discretization when the underlying model is B-spline of order $0$, which is, in general, too coarse an approximation subspace. In this paper, we proposed a CS framework for single-pixel imaging that uses an arbitrary SI signal model and showed that B-splines of higher orders lead to efficient implementations and better reconstruction results in comparison to the conventional discretization setting. Furthermore, directly recovered B-spline signal representation is an alternative approach whose polynomial interpretation offers many practical advantages in image processing applications.

The fact that the B-splines correspond to scaling functions in spline-wavelet frames has led us to use the biorthogonal wavelets as sparsity-inducing dictionaries. We proved that the coefficients recovered by solving the CS optimization problem are wavelet coefficients corresponding to the orthogonal projections of the signal onto the continuous-domain analysis wavelet functions when the B-spline representation function is equal to the scaling function. However, the extensive experimental results have shown that, in general, higher orders of B-spline representation functions yield better reconstructions even when the scaling function corresponds to a B-spline of a low order. That is, it is more important to choose a proper underlying signal model than to match the representation function with the scaling function of the frame. The discretization remains exact and the sequence of representation coefficients are assumed to be sparse in a discrete transform basis.

We offered an efficient matrix-free implementation of the proposed framework and conducted it on synthetic and real-world measurement data. The experiments have shown that the proposed discretization method overachieves the conventional one in both noise-free synthetic and noisy real-world scenarios.

\appendix
\section{} 
\label{sec:appendixA}

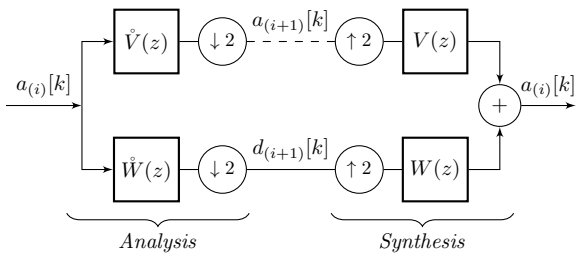
\begin{figure}[h]
    \centering
    \begin{tikzpicture}[auto, node distance=1cm, >=latex', scale=0.8, every node/.style={scale=0.8}]
        \node [input, name=x_input] (x_input) {};
        \node [input, right of=x_input, name=cross, node distance = 1.25cm] (cross) {};
        \draw [->] (x_input) -- node{$a_{(i)}[k]$} (cross);
        \node [block, below right of=cross, node distance=1.5cm] (FilterH) {$\mathring{W}(z)$};
        \draw [->] (cross) |- node{}(FilterH);
        \node [sum, right of=FilterH, scale=0.9, node distance = 1.25cm] (downsampleH){$ \downarrow 2$};
        \draw [-] (FilterH) -- node{}(downsampleH);
        \node [input, right of=downsampleH, name=crossH, node distance = 0.75cm] (crossH) {};
        \node [block, above right of=cross, node distance=1.5cm] (FilterL) {$\mathring{V}(z)$};
        \draw [->] (cross) |- node{}(FilterL);
        \node [sum, right of=FilterL, scale=0.9, node distance = 1.25cm] (downsampleL){$\downarrow 2$};
        \draw [-] (FilterL) -- node{}(downsampleL);
        \node [input, right of=downsampleL, name=crossL, node distance = 0.75cm] (crossL) {};
        \node [sum, right of=downsampleH, scale=0.9, node distance = 2.25cm] (upsampleHrec){$\uparrow 2$};
        \node [block, right of=upsampleHrec, node distance=1.25cm] (FilterHrec) {$W(z)$};
        \draw [-] (upsampleHrec) -- node{}(FilterHrec);
         \node [sum, right of=downsampleL, scale=0.9, node distance = 2.25cm] (upsampleLrec){$\uparrow 2$};
        \node [block, right of=upsampleLrec, node distance=1.25cm] (FilterLrec) {$V(z)$};
        \draw [-] (upsampleLrec) -- node{}(FilterLrec);
        \draw [-, dashed] (downsampleL) -- node{$a_{(i+1)}[k]$}(upsampleLrec);
        \draw [-] (downsampleH) -- node{$d_{(i+1)}[k]$}(upsampleHrec);
        \node [sum, above right of=FilterHrec, name=crossOut, node distance = 1.5cm] (crossOut) {$\suma$};
        \draw[->] (FilterHrec) -| node{}(crossOut);
        \draw[->] (FilterLrec) -| node{}(crossOut);
        \node [output, name=out, right of=crossOut, node distance = 1.25cm] (out) {};
        \draw[->] (crossOut) -- node{$a_{(i)}[k]$}(out);
        
        \draw [decorate,decoration={brace,amplitude=5pt,mirror, raise=1.5ex}] (1,-1.5) -- (4,-1.5) node[midway,yshift=-3em]{\textit{Analysis}};
        \draw [decorate,decoration={brace,amplitude=5pt,mirror, raise=1.5ex}] (5.35,-1.5) -- (8.35,-1.5) node[midway,yshift=-3em]{\textit{Synthesis}};
    \end{tikzpicture}
    \caption{\textbf{Perfect reconstruction wavelet filter bank.}}
    \label{fig:WaveletFilterBank}
    \vspace{-1.75ex}
\end{figure}

Wavelet forward and inverse transforms can be accomplished efficiently by using a perfect reconstruction filter bank described by the block diagram in Figure \ref{fig:WaveletFilterBank}, 
which is an extension of Mallat's fast wavelet algorithm \cite{Mallat:MRA} for non-orthogonal basis functions \cite{Vetterli:BiorWavelets}. We denote by $\mathring{V}(z)$ and $\mathring{W}(z)$ the $z$-transforms of the discrete-time filters in the filter bank associated with the analysis scaling and wavelet functions $\mathring{\varphi}(t)$ and $\mathring{\psi}(t)$, respectively. Analogously, $V(z)$ and $W(z)$ are the $z$-transform of the discrete-time filters associated with the synthesis (reconstruction) scaling and wavelet functions $\varphi(t)$ and $\psi(t)$, respectively.

\section{} 
\label{sec:appendixB}
\subsection{}
\label{subsec:appendixB1}
By using the signal form in \eqref{eq:SignalinSIspace} and inner product definition in \eqref{eq:L2innerProduct}, the expansion of \eqref{eq:SamplingInSIdef} is given by
\begin{align*}
    c[k]&= \left\langle  \sum_{l \in \mathbb{Z}} a_{(0)}[l]\varphi(t-l),\zeta(t-k)  \right\rangle \nonumber \\ 
        &= \sum_{l \in \mathbb{Z}} a_{(0)}[l] \left\langle\varphi(t-l),\zeta(t-k) \right\rangle \\
        &= \sum_{l \in \mathbb{Z}} a_{(0)}[l] r[k-l] = a_{(0)} * r[k]. \nonumber
\end{align*}

\subsection{}
\label{subsec:appendixB2}
In the general SI sampling framework, to obtain the approximation coefficients at resolution $2^0$, $f(u,v)$ is filtered with an analog prefilter and uniformly sampled. Samples $c[k,l]$ for a separable two-dimensional sampling kernel $\zeta(u,v)$ and a signal generator $\varphi(u,v)$ are given by
\begin{align}
    \label{eq:app_SamplesDef}
    c[k,l] &= \int \limits_{-\infty}^{\infty} \int \limits_{-\infty}^{\infty}f(u,v) \zeta(u-k,v-l)dudv \\
           &= \int \limits_{-\infty}^{\infty} \int \limits_{-\infty}^{\infty} \left(  \sum_{m\in \mathbb{Z}} \sum_{n\in \mathbb{Z}} a_{(0)}[m,n] \varphi(u-m) \varphi(v-n) \right. \nonumber \\
           & \qquad \qquad \qquad \qquad \qquad \quad \times \zeta(u-k) \zeta(v-l) \Bigg)dudv \nonumber \\
           &= \sum_{m\in \mathbb{Z}} \sum_{n\in \mathbb{Z}} a_{(0)}[m,n] \int \limits_{-\infty}^{\infty} \varphi(u-m) \zeta(u-k)  du \nonumber \\
           & \qquad \qquad \qquad \qquad \qquad \quad \times \int \limits_{-\infty}^{\infty} \varphi(v-n) \zeta(v-l)dv. \nonumber
\end{align}
By using the inner product definition in \eqref{eq:L2innerProduct}, \eqref{eq:app_SamplesDef} becomes
\begin{align*}
    c[k,l] &=  \sum_{m\in \mathbb{Z}} \sum_{n\in \mathbb{Z}} a_{(0)}[m,n] \left\langle \varphi(u-m), \zeta(u-k) \right\rangle \\
           & \qquad \qquad \qquad \qquad \qquad \qquad \ \ \times  \left\langle  \varphi(v-n), \zeta(v-l)  \right\rangle \nonumber \\
           &= \sum_{m\in \mathbb{Z}} \sum_{n\in \mathbb{Z}} a_{(0)}[m,n] r[k-m] r[l-n]. \nonumber
\end{align*}
The inner products of the basis functions are the sampled cross-correlation sequences defined in \eqref{eq:CrossCorrelationSeq}. Thus, the samples $c[k,l]$ are given by the separable two-dimensional convolution between the expansion coefficients $a_{(0)}[k,l]$ and the sampled cross-correlation sequence $r[k]$.

\section*{Acknowledgments}
The authors would like to thank Ivan Ralašić for providing codes for the off-the-shelf compressive imaging system and fruitful discussion on the system parameters and geometry, and to Professor Sven Lončarić for lending an industrial camera and a projector for the experiments. The authors are also grateful to the anonymous reviewers for their valuable comments and suggestions, which immensely helped improving the quality of the paper.

\section*{Funding}
This research was supported by the European Regional Development Fund [KK.01.1.1.01.0009] (DATACROSS); and the Croatian Science Foundation [IP-2019-04-6703].

\balance

\bibliographystyle{elsarticle-num}
\section*{\refname}
\bibliography{Bibliography_Template}

\end{document}